\documentclass[draftcls, onecolumn]{IEEEtran}

\usepackage{mathtools}
\usepackage[latin1]{inputenc} 
\usepackage{soul} 
\usepackage{multirow} 
\usepackage{float}
\usepackage{algorithm}
\usepackage{color}
\usepackage{amsmath,amssymb,amsfonts,amsthm}
\usepackage{algorithmicx}
\usepackage{algpseudocode}
\usepackage[caption=false]{subfig}
\usepackage[latin1]{inputenc}
\usepackage{array}
\usepackage{graphicx}

\begin{document}
	\title{Heuristic Chaotic Hurricane-aided Efficient Power Assignment for Elastic Optical Networks}
	\author{Layhon~R.~R.~dos~Santos, ~Taufik~Abr\~ao
\thanks{Taufik~Abr\~ao is with the Department of Electrical Engineering, Londrina State University, Parana, Brazil e-mail: {taufik@uel.br}.}
\thanks{Layhon~R.~R.~dos~Santos are with the Department of Electrical Engineering, Technological Federal University, Corn\'elio Proc\'opio, Parana, Brazil e-mail: {lay.lyns@hotmail.com}}
}
	
\maketitle
\begin{abstract}
In this paper we propose a dynamical transmission power allocation for elastic optical networks (EONs) based on the evolutionary hurricane search optimization (HSO) algorithm with a chaotic logistic map diversification strategy with the purpose of improving the capability to escape from local optima, namely CHSO. The aiming is the dynamical control of the transmitted optical powers according to the each link state variations due to traffic fluctuations, channel impairments, as well as other channel-power coupling effects. Such realistic EON scenarios are affected mainly by the channel estimation inaccuracy, channel ageing and power fluctuations.  The link state is based on the channel estimation and quality of transmission (QoT) parameters obtained from the optical performance monitors (OPMs). Numerical results have demonstrated the effectiveness of the CHSO to dynamically mitigate the power penalty under real measurements conditions with uncertainties and noise, as well as when perturbations in the optical transmit powers are considered.
\end{abstract}
	

\section{Introduction}	
The  growth of the traffic demand with heterogeneous characteristics associated to the increment of the SNR rate requirements has pressing the development of dynamical optical networks. Currently, the technological maturity of devices, equipment and protocols provides the use of dynamical flexible grid-rate elastic optical network (EON). In the EONs, the lightpaths with adjustable bandwidth, modulation level and spectrum assignment can be established according to actual traffic demands and quality of service (QoS) requirements~\cite{ref:1}-~\cite{ref:3}. In addition, the quality of transmission (QoT) of each lightpath is evaluated previously to resources allocation purpose, as well as to obtain reliable optical connectivity~\cite{ref:3}\cite{ref:4}. The best knowledge of the QoT is needed in the design and operation phases, owing to the margin has to be added in the network when the QoT is not well established~\cite{ref:2}. The QoT prediction can utilizes different methodologies based on sophisticated analytical models, approximated formulas and optical performance monitors (OPMs)~\cite{ref:2}\cite{ref:5}. The QoT estimation with OPMs distributed in the route or in the coherent receiver can be appropriated in term of precision and computational complexity when integrated in to the active control plane to provide the link conditions in real time~\cite{ref:1}\cite{ref:4}. However, it is important to consider the limited accuracy of the OPMs that increase the measurements uncertainty considering the channel impairments (including linear and nonlinear effects), receiver architecture and noise, which decrease the performance of the channel state estimation~\cite{ref:5}~\cite{ref:6}. In addition, the power dynamics related to the channel-power coupling effects, which are influenced by the network topology, traffic variation, physics of optical amplifiers and the dynamic addition and removal of lightpaths can cause optical channel power instability and result in QoT degradation~\cite{ref:1}. Moreover, the interactions between lighpaths in some routes of the network can generate fluctuations to form closed loops and create disruptions.   
	
The power, routing, modulation level and spectrum assignment (PRMS) is usually determinate in the planning stage of the network and margins are included considering the QoT inaccuracies, equipment ageing, inter-channel interference, as well as uncertainties of the optical power dynamics~\cite{ref:7}-\cite{ref:12}. However, there are some investigations to development of resource allocation algorithms based on OPMs with reduced margins, which have considered ageing and inter-channel interference to configurable transponders with launch powers~\cite{ref:5}, regenerator placement~\cite{ref:2} and the optimization of the physical topology for power minimization~\cite{ref:11}. These algorithms can be based on derivative-free optimization (DFO), constrained direct-search algorithms~\cite{ref:5}, and mixed integer linear programming (MILP)~~\cite{ref:11}. Furthermore, in~\cite{ref:6} an adaptive proportional-integral-derivative (PID) with gains auto-tuning based on particle swarm optimization (PSO) to dynamically controls the transmitted power according to the OPMs measurements for mixed line rate (MLR) was proposed. Previous investigations for legacy single rate network for power control adjustment to the optical-signal-to-noise-ratio (OSNR) optimization considering the physical impairments were conducted based on a game-theory-based~\cite{ReF:23} and (PID) back propagation (BP) neural networks~\cite{ref:14}.  Moreover, the power allocation optimization aiming at obtain energy-efficient optical CDMA systems using different programming methods is carried out in  \cite{PENDEZA_2019}. Such optimization methods including augmented Lagrangian method (ALM), sequential quadratic programming method (SQP), majoration-minimization (MaMi) approach, as well as Dinkelbach's method (DK) were compared under the perspective of performance-complexity tradeoff.   The findings reported in the previous papers assume that there are no impact of queuing issues on the optical network convergence and performance.  To highlight this important aspect, in \cite{KUMAR_2017} the authors carried out a review on the role of the queuing theory-based statistical models in wireless and  optical networks. 

The contributions of this work include: a) proposing an effective power-efficient assignment strategy based on heuristic chaotic hurricane-aided approach; b) investigating systematically the CHSO input parameter optimization aiming at improving the performance-complexity tradeoff of the proposed algorithm; c) validating the proposed power allocation method for different realistic EON channel conditions, {i.e.}, non-perfect monitoring of the OPMs, channel ageing effects, dynamical scenarios, including power instability,  particularly in EONs.

\begin{figure}[htbp!]
	\centering
	\includegraphics[trim={0mm 0mm 0mm 0mm},clip, width=1\linewidth]{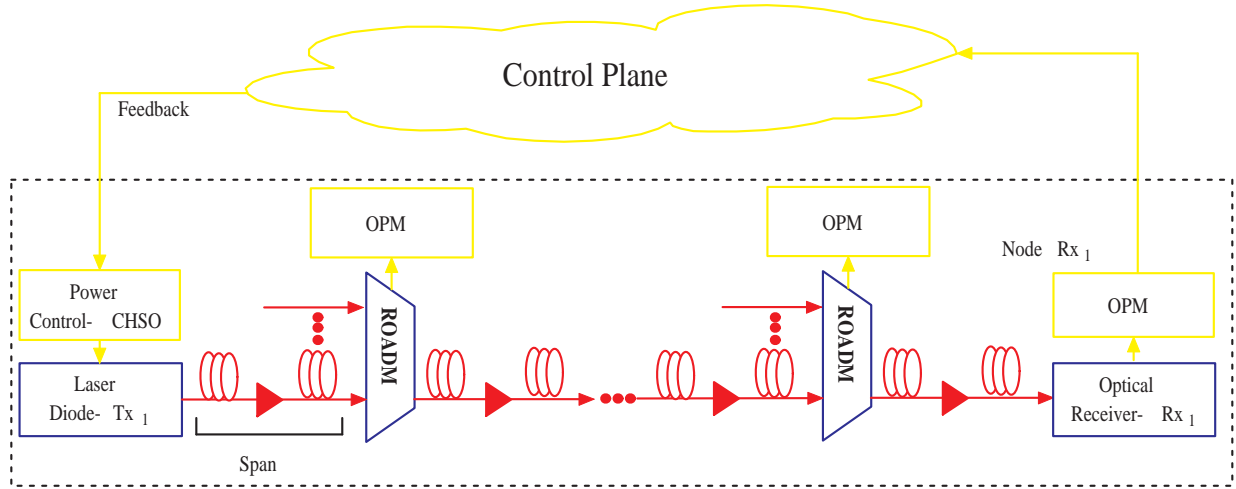}
	\vspace{-5mm}
	\caption{Elastic optical network topology highlighting the chaos heuristic-based power control block composed by a chaotic hurricane search optimization (CHSO) scheme based  on QoT estimation.}
	\label{fig:1}
\end{figure}

The channel estimation in terms of QoT parameters obtained from the OPMs and deployed in the algorithm updating is illustrated in Fig.~\ref{fig:1}. 
The proposed scheme for the EONs power allocation is named hereafter the chaotic-hurricane search optimization (CHSO). Hence, normalized mean square error (NMSE), convergence, power penalty, probability of success and computational complexity are evaluated aiming to corroborate the effectiveness and efficiency of the proposed resource allocation strategy, specifically operating in EONs. Moreover, comparisons have been performed assuming a convex optimization through the gradient descent (GD) \cite{ref:15, ReF:36}. Finally, power-efficient assignment by CHSO decrease the margins, improve the energy efficiency (EE), reduce the costs while attains a better performance-complexity tradeoff.

\section{Proposed Scheme}

The proposed scheme utilizes the information collected from the OPMs to control dynamically the level of the launch power of the lightpaths. The power adjustment considers the QoT inaccuracies, equipment ageing, inter-channel interference, as well as the variation of the optical power dynamics. Differently of the others approaches based on intelligent systems~\cite{ref:15}, it is not necessary the training phase and the proposed scheme can be performed in near-real time. In addition, the power budget is determinate in the planning stage of the network and margins are included~\cite{ref:3}~\cite{ref:2} and the proposed scheme will act during the regular operation of the EON. For the proposed scheme it is considered that the lightpaths were previously established from the resource allocation algorithms associated with route, modulation, bandwidth and spectrum.  

The proposed scheme continuously update the transmitter launch power for each lightpath in response to dynamic OPMs information, it is considered a communication delay between the OPMs in the receiver node, control plane and transmitter adjustment. This process can encompass the delay related to the duration of the OSNR estimation in the OPM, the control message transmission duration, the processing time, the actuation phase in the transmitter and the round-trip delay. In this sense, considering the current technology, each algorithm updating can be estimated in 100 ms or less~\cite{ref:2}. Therefore, the time needed to close the loop related to the signal latency and the other operations needed for controlling the transmitted power is assured. 

The proposed power allocation algorithm utilizes the chaotic hurricane, which is a new hybrid algorithm based on the hurricane search optimization (HSO) associated with probability distribution from the chaotic maps instead of uniform distribution of the traditional HSO. The objective is obtain an algorithm with balance between the exploration (diversification) and the exploitation (intensification) to improve the algorithm capability to escaping of the local solutions and the amelioration of the velocity of the convergence without affecting the quality of the algorithm solutions~\cite{ref:16}-\cite{ReF:103}. 

The HSO is a metaheuristic algorithm for global optimization considering single-objective~\cite{ReF:17} and multi-objective~\cite{ReF:101} optimization problems, inspired by natural phenomena on the hurricanes behavior, where wind parcels move in a spiral course moving away from a low-pressure zone called the eye of the hurricane. These wind parcels search for possible new eye position, which represents a lower pressure zone to find out the optimal solution. The performance is very competitive compared to others metaheuristics optimization algorithms, such as gravitational search algorithm (GSA) and particle swarm optimization (PSO). Although there is a variety of optimization algorithms, the development of new optimization algorithms have been motivated by the no free lunch (NFL) theorems for optimization, which have proved that an universally efficient optimization algorithm does not exist. Moreover, the particularities and characteristics of the optimization problem strongly affects the capacity of the optimization algorithm to finding the optimal solution in global optimization problems~\cite{ReF:17}. Herein, it is important investigate several distinct optimization algorithms for different optimization problems considering the related aspects. In addition, the application of chaos theory alone or jointly with other algorithms such as ant colony algorithm (ACO)~\cite{ref:16}, firefly algorithm (FA) \cite{ref:170} and PSO~\cite{ReF:103} have improved the optimization algorithms. Chaos presents a non-repetitive nature that increase the random search characteristics of the optimization methods, as well as increases the ability to get away from local solutions. In general, chaotic maps based on the complex behavior of a nonlinear deterministic system are utilized to optimization goal.

\section{Performance Evaluation}
The EON physical layer is composed by transmitters with adjustable modulation format, SNR rate and level of launch power, an erbium-doped fiber amplifier (EDFA) per span, ROADMs and receivers with digital signal processing capability to compensate the dispersion effects. The ROADMs present equalization to compensate undesired spectrum tilting due to EDFAs. In addition, the EDFAs operated in an automatic gain controlled (AGC) mode according to each ROADM to achieve spectral tilt correction. The lightpaths are represented as Nyquist wavelength division multiplexing (WDM) superchannels with bandwidth 
\begin{equation}
\Delta f_i = \frac{\xi_i}{c_i}, \quad i=1, \ldots, M
\end{equation}
where $\xi_i$ is the traffic demand data rate (Gbps) of the $i$th channel, $M$ is the number of channels and $c_i$ is the spectral {efficiency} defined by modulation format of the $i$th channel, Table~\ref{tab:1}. The  ${\rm SNR^*_{\rm B2B}} (c_i)$ is the back-to-back signal-to-noise ratio target for the $i$th channel required to achieve error-free considering forward-error-correction (FEC) codes with bit-error-rate requirement of ${\rm BER}^* = 4\cdot 10^{-3}$.

\begin{table}[!htbp]
	\centering
	\caption{Modulation format, spectral efficiency and SNR}
	\small
	\begin{tabular}{ccc}	
		\hline
		\multirow{2}{*}{Modulation format} & Spectral Efficiency  & SNR$_{\rm B2B}^{{*}}$ \\
		& (bps/Hz) & (dB) \\ \hline
		PM-BPSK           & 2        & 5.50     \\
		PM-QPSK           & 4        & 8.50     \\
		PM-8QAM           & 6        & 12.50    \\
		PM-16QAM          & 8        & 15.15    \\
		PM-32QAM          & 10       & 18.15    \\
		PM-64QAM          & 12       & 21.10  \\
		\hline
	\end{tabular}
	\label{tab:1}
\end{table}

Hence, to obtain an appropriate QoT, the {effective back-to-back signal-to-noise ratio for the channel $i$ (${\rm SNR}_{{\rm B2B},i}$) must be ${\rm SNR}_{{\rm B2B},i} \geq {\rm SNR_{B2B}^*} (c_i)$}. This formulation can be defined as a problem of residual margin (RM). The RM in the  $i$th channel can be defined as~\cite{ref:15}:
\begin{equation}\label{eq:5}
\Psi_i= \frac{{\rm SNR}_{{\rm B2B},i}}{{\rm SNR}_{\rm B2B}^* (c_i)},
\end{equation}
while the RM in vector form is represented as ${\bf \Psi}=[\Psi_1,\Psi_2,\cdots,\Psi_M ]^{1\times M}$. The concept of residual margin ${\bf \Psi}$ in WDM systems can be formulated as an optimization problem, which the objective is to minimize the RM of all the $M$ WDM channels in the sense of ${\bf \Psi}^*=[1,\cdots,1]^{1\times M}$, while guaranteeing the QoT. 

Such RM optimization problem can assume an optical network topology as static in a short-time due to the multi-stage traffic demand. Therefore, the optimization problem reduces to the efficient power assignment problem~\cite{ref:5}:
\begin{align} \label{eq:6}
\mathop{\text{minimize}}\limits_{{\bf p} \in\mathbb{R}^M} \quad & J({\bf p})=\sum_{i=1}^M p_i\,\, = \,\, {\bf 1}^T {\bf p}\notag \\
\text{s.t.}\quad \text{ \small (C.1)} &\quad \displaystyle {{\rm SNR}_{{\rm B2B},i}} \geq {\rm SNR}_{\rm B2B}^*(c_i)\\
\text{ \small (C.2)}& \quad \displaystyle  {\xi_i}  \geq {\rm \xi}_i^{\rm min} \notag\\
\text{ \small (C.3)} &  \quad \displaystyle  {p}_{\rm min}  \leq p_i \leq {p}_{\rm max} \notag
\end{align}
where ${\bf p}=[p_1,p_2,\cdots, p_M ]^T$ is the optical power vector and $p_i$ the transmitted power for the $i${\rm th} lightpath, subject to the constraints related to power budget and SNR required to achieve QoT \cite{ref:4}; moreover, $\xi_i^{\rm min}$ is the SNR rate supported transmission, $p_{\rm min}$  and $p_{\rm max}$ is the minimum and maximum value considered as allowable transmitted power, respectively. 

The quality of the RM optimization attained by different methods can be evaluated via the Euclidean distance between ${\bf \Psi}$ and ${\bf \Psi}^*$. Mathematically, this is expressed as: 	
\begin{align} \label{eq:7}
& \mathop{\text{minimize}}\limits_{{\bf p} \in\mathbb{R}^M} \quad J_1({\bf p})=\|{\bf \Psi}^* - {\bf \Psi}\|_2 \notag \\
& \text{s.t.}\quad \text{ \small (C.1), (C.2), (C.3) }\, \text{of eq.~\eqref{eq:6}}.
\end{align}

From eq.~\eqref{eq:6}, the problem in the original form is not convex. Therefore, it do not guarantee the minimum global of ${\bf \Psi}$ applying nonlinear programming (NLP), because can to result in a solution that is far from global solution ${\bf \Psi}^*$. In this sense, heuristic evolutionary optimization methods, such as the HSO, have the advantage of achieving in a polinomial processing time an high quality solution, which is not necessary the optimum solution.  For comparison purpose, in the numerical results it will be evaluated descent gradient (GD) proposed~{\cite{ReF:36}} with convex formulation~{\cite{ref:15}}. The first constraint from eq.~\eqref{eq:6} is based on a Gaussian noise (GN) model to establishment the QoT in the lightpath~\cite{ReF:20}~\cite{ReF:20a}. The $i$th lightpath route is originated in the source $s_i$ and destination $d_i$ traversing the number of span of $N_{s_{i}}$, considering the number of spans shared with interfering $j$th lightpaths $N_{s_{ij}}$. 

For the proposed evaluation scenario in Fig. \ref{fig:1}, it is considered that the $i$th lightpath characteristics such as modulation formats, routes, and spectral orderings of all the connections were previously determined, and it includes the design margin~$M_d(\tau)$ due to the QoT model inaccuracies and ageing margin of the transponder $M_t(\tau)$ on its sensitivity, modeled as a function of time $\tau$. Hence, the ${\rm SNR}_{{\rm B2B},i}$ for the $i$th channel can be defined by:
\begin{equation}
{\rm SNR}_{{\rm B2B},i}(\tau) = {\rm SNR}_i (\tau) - M_d(\tau) - M_t(\tau)
\label{SNR_tau}
\end{equation}
where the parameters from eq.~\eqref{SNR_tau} can be modeled as linear or nonlinear functions of time $\tau$~\cite{ref:2}. Herein, we have adopted the following linear function of $\tau$:
\begin{equation}\label{eq:8}
\displaystyle M_t(\tau)=\frac{M_t(\tau_{\rm end})-M_t(\tau_0)}{\Delta \tau} \cdot \tau
\end{equation}	
where $M_t(\tau_{\rm end})$ and $M_t(\tau_0)$ are the transponder margin for  End-of-Life (EoL) and  Begin-of-Life (BoL) time, respectively, while $\Delta \tau = \tau_{\rm end}- \tau_{0}$ is the network's lifetime.  Besides, the first term ${\rm SNR}_i(\tau)$ assumes GN model, while includes the linear and nonlinear noise effects for the $i$th channel \cite{ReF:20}:
\begin{equation}
\label{eq:8}
{\rm SNR}_i(\tau)= \frac{p_i}{\left[G^{\rm ASE}_i(\tau)+G^{\rm SCI}_i(\tau)+G^{\rm XCI}_i(\tau)\right] \Delta f_i}
\end{equation}
where $G_i$ is the power spectral density (PSD) in [W/Hz] of the $i$th channel, $G_i^{\rm ASE}(\tau)$ is the amplified spontaneous noise (ASE) noise, $G_i^{\rm SCI}(\tau)$ is the self-channel interference (SCI), and $G_i^{\rm XCI}(\tau)$ the cross-channel interference (XCI) for the $i$th channel.  

The PSD of the ASE noise is given by \cite{ref:2}\cite{ReF:20}:
\begin{equation}
\label{eq:9}
G_i^{\textsc{ase}}(\tau)= h   v   N_e  \left(\sum_{n_r=1}^{N^{\textsc{roadm}}_i} (A^{\textsc{roadm}}_{e,i}-1) + \sum_{e=1}^{N^{\rm span}_{i}} (A^{\rm span}_{e,i}-1)\right)
\end{equation}
where $h$ is the Planck's constant, $v$ is the carrier frequency,  $N_e$ is the noise figure of the \textsc{edfa}, $N^{\textsc{roadm}}_i$ and $N^{\rm span}_i$ are respectively the span and \textsc{roadm} number of the $i$th user. $A^{\textsc{roadm}}_{e,i} (\tau)$ and $A^{\rm span}_{e,i}(\tau)$ are the losses of the $e$th span and $e$th ROADM from the $i$th user, respectively, being the last given by:
\begin{equation}
\label{eq:9a}
A^{\rm span}_{e,i}(\tau)= L_{e,i} \cdot \alpha_{\rm loss} (\tau) + c_{e,i} \cdot c_{\rm loss} (\tau) + s_{e,i} \cdot s_{\rm loss} (\tau)
\end{equation}
where for the $e$th span of the $i$th user, the $L_{e,i}$ is span length; $c_{e,i}$ is connection number, and $s_{e,i}$ is the splice number; while $\alpha_{\rm loss}$, $c_{\rm loss} (\tau)$ and  $s_{\rm loss} (\tau)$ can be modeled as functions of time, representing the fiber attenuation, the connector's loss and the splice loss, respectively~\cite{ref:2}.

The PSD of the SCI noise is given by:
\begin{equation}\label{eq:10}
G_i^{\rm SCI}(\tau)= \frac{3\gamma^2}{2 \pi \alpha |\beta_2|} \sinh^{-1} \left( \frac {\pi^2 |\beta_2|} {2\alpha}\Delta f_i^2\right)G_i^3 N_{s_{i}}
\end{equation}
being $\gamma$ the nonlinear parameter and $\beta_2$ is the is the group velocity dispersion.

The PSD of the XCI noise is given by~\cite{ReF:20},
\begin{equation}
\label{eq:11}
G_i^{\rm XCI}(\tau)= \frac{6\gamma^2}{\alpha^2} G_i \sum_{j \neq i} \frac{\alpha}{4 \pi |\beta_2|} {\rm log} \left| \frac{|f_i-f_j|+\Delta f_j/2}{|f_i-f_j|-\Delta f_j/2} \right| G_j^2 N_{s_{i}}
\end{equation}
being $G_j$ the PSD of $j$th interfering channels. Therefore, can be obtained a bit error rate (BER) expressed as a function of the SNR, which takes into account the baud-rate, FECs limit BER and the modulation format of the $i$th channel~\cite{ReF:22}~\cite{ReF:22a}, as follows:
\begin{equation}
{\rm BER}_i = \vartheta ({\rm SNR}_{{\rm B2B},i})
\label{Pb}
\end{equation}
where the function $\vartheta (\cdot)$ is defined by modulation format~\cite{ref:2}. 

The QoT prediction consists of developing  a systematic procedure for the evolution of the vector ${\bf p}$ in order to reach the optimum value ${\bf p}^*$, based on the ${\rm SNR}_i$, ${\rm SNR}_{\rm B2B}(c_i)$, ${\rm BER}_i$, ${\rm BER}^*_i$ values. Theses values are monitored by OPMs at add, through and drop node by channel estimation and reported to the control plane to guarantee the QoT. The channel estimation quality is affected by three main assumptions: 
\begin{enumerate}
	\item  non-perfect monitoring of the OPMs considering their limited accuracy due to channel impairments (linear and nonlinear effects) and the receiver architecture, as well the noise measurement and peaks occurrence caused by polarization mode dispersion (PMD) effects \cite{ref:3,ref:2,ref:6,ReF:21,ReF:38}. Theses uncertainties can be modeled as a random variable $\delta_i$ added to the ${\rm SNR}_i$,which follows a Log-Normal distribution $\mathcal{LN}\displaystyle (\mu, \sigma)$. Therefore, the estimated $ {\rm SNR}_i $ can be modeled by~\cite{ref:6}:
\begin{equation}
\begin{array}{c}
\widehat{\rm SNR}_i= {\rm SNR}_i (1+\epsilon_i) \quad \forall_i {, \,\,\,} 
\epsilon_i  \sim \mathcal{LN} \displaystyle (\mu, \sigma) 
\end{array}
\end{equation} 
\item ageing resulting from increases fiber losses due to splices to repair fiber cut, detuning of the lasers leading to misalignment with optical filters in the intermediate and add/drop nodes. These values can be modeled by eqs.~\eqref{eq:9}-\eqref{eq:9a} as function of time $\tau$, assuming the parameter values based on Begin-of-life (BoL) and End-of-life (EoL) in an elastic optical network. 
\item power instability resulting from power variations due linear and nonlinear effects associated to the optical fiber and coupling, both influenced by traffic variation, network topology, physic aspects of the EDFA and ROADM at add/drop channel, and unpredictability of fast time-varying penalties, such as polarization effects. Theses values can be modeled as a power perturbation in the input power of the $i$th user: 
\begin{equation}\label{eq:power_inst}
p_i^{\circ}[n]=  {\rm pert}[n] + 10\log_{10} \frac{p_i}{10^{-3}}  \qquad \rm [dBm]
\end{equation}
where the  power perturbation function is modeled as 
\begin{equation}\label{eq:power_inst2}
{\rm pert}[n] = {A^n \cdot {\rm sin} (n\pi/2)},
\end{equation}
with $A^n$ being the peak of the perturbation in [dB], $n$ is a discrete-time index, and $p_i$ the nominal transmitted power for the $i$th lightpath. This model assumes power fluctuations propagation across the network nodes~\cite{ref:6}.
\end{enumerate}

The full knowledge of the QoT parameters during the estimation of the $i$th channel increases reliability and  enables design solutions considering the {$\bf p^*$} for different bit rates requirement in the lightpath. In this sense, mixed line rates (MLR) networks have focused on optimum launch power, obtaining suitable cost minimization~\cite{ReF:26}, combined to the maximization of the number of established connections~\cite{ReF:23}, while reducing the transponder cost~\cite{ReF:24} and improving the launch power versus regenerator placement tradeoff~\cite{ReF:25}. However, when the QoT parameters for the $i$th channel is not  known perfectly, power penalty (PP) occurs, being modeled as:
\begin{equation}\label{eq:13}
\displaystyle  \bar{p}_i(n) =  10 \cdot {\log} \left( \frac{p_i(n)}{p_i^*} \right) \qquad [dB]
\end{equation}
where $p_i (n)$ is the launch power at the $n$th iteration evaluated by the proposed scheme and $\displaystyle p_i^*$ is the optimal launch power obtained in the static planning phase. The $\displaystyle p_i^*$ value is defined considering the perfect knowledge of the QoT parameters~\cite{ref:5}. Negative values of $\bar{p}_i$, {i.e}, $\bar{p}_i^{-}$, mean that the measured BER did not reach the  BER$^*$, while positive values  ($\bar{p}_i^{+}$) mean that the BER$^*$ is reached with expenditure energy. Therefore, for availability of the $i$th lightpath, the margins ($m_i$) should satisfying the condition of $m_i\geq {\bar{p}_i^-}$.

In context of margins, in~\cite{ref:13} a system margin (SM) is adjusted by a ML based on the maximum-likelihood principles to improve the QoT prediction of new lightpaths. The predict parameters can provided more accurate QoT of not-already-established lightpaths compared to the limited amount of information available at the time of offline system design. In~\cite{ref:14} is proposed a ML-based classifier to predict if the candidate lightpath presents suitable bit error rate (BER) considering the traffic volume, modulation format, lightpath total length, length of its longest link, and number of lightpath links. To train of the ML classifier is based on the OPMs or in the BER simulation, which is utilized  in the absence of real field data. In~\cite{ref:15} is performed the optimization of transmitted power to maximize minimum margin and to maximize a continuously variable data rate. The Gaussian noise nonlinearity model is utilized to expresses the SNR in each channel as a convex function of the channel powers. Convex optimization is performed with objectives of maximizing the minimum channel margin.

Therefore, the progress in the network planning, design and active operation control has become margins an important resource to be optimized \cite{ref:3}\cite{ref:2}. In this sense, the margin in each lightpath should be as little as possible to ensure guarantee reliable optical connectivity. The reducing of the excess margin can be utilized to increase the maximum transmission distance, reduce the number of regenerators, as well as postpone the installation of more robust transponders than are closely necessary in the beginning of the network operation~\cite{ReF:21}. Several efforts have been made to become the margins variable and adjustable to increase the network capacity and decrease the costs of the network implantation and operation~\cite{ref:1}-~\cite{ref:6}. In this sense, the determination of the level of transmitted power is performed in the planning stage of the network and a SM is included considering the uncertainties of the OPMs measurements and optical power dynamics~\cite{ref:7}\cite{ReF:20}.

\section{Adaptive-Chaotic Hurricane Search Optimization}
In the HSO, the eye (lower pressure zone) is related to the best solution of the hurricane structure and can be represented at $n$th iteration by the matrix ${\bf P}[n] = [{\bf p}_{1} [n] \,\, {\bf p}_{2} [n] \cdots \,\, {\bf p}_{K} [n]]^{\rm T} \in \Re^{K\times M}$, which is composed by $K$ wind parcels, defined as ${\bf p}_k[n]= [{p}_{k,1}[n], \,\, {p}_{k,2}[n], \cdots, \,\, {p}_{k,M}[n]]\in \Re^{1\times M}$, while  the hurricane eye is the best candidate vector solution at $n$th iteration, written as $\hat{\bf p}[n]=[\hat{p}_1[n],\,\,\hat{p}_2[n], \,\,\cdots,\,\,\hat{p}_M[n]]\in \Re^{1\times M}$. Besides, the parameter $K$ is composed by wind parcels factor $N_w$ and $M$ channels, resulting $K=M\cdot N_w$.  The pressure function at the $n$th iteration for the hurricane eye $p_{\bf \hat{p}}[n]$, as well as for the candidate solutions {$p_{{\bf P}_k}[n]$} is measured by athe fitness function  in eq.~\eqref{eq:7}, {\it i.e}.,$ {\rm pressure}({\bf \hat{p}}[n])= \upsilon  J_1({\bf \hat{p}}[n])$, where $\upsilon$ is a constant.

The $k$th wind parcel on the $n$th iteration moves around the eye according to:
\begin{equation}\label{eq:1}
{r_{k}[n]}(\theta_{k}[n])={r_0} \cdot \exp{(z_{k}[n] \cdot \theta_{k}[n])},
\end{equation}
where ${r_{k}[n]}$ and $\theta_{k}[n]$ are respectively the radial and angular coordinate of the power increasing of the {$k$th} wind parcel at the {$n$th} iteration. The variable ${r_{0}}$ represents the initial value of $r_{k}[n]$ and From eq.~\eqref{eq:1}, the variable $z_{k}[n]$ is the rate of the increase of the spiral at $n$th iteration. Indeed, the behavior of the $k$th wind parcel in the $n$th iteration follows a logarithmic spiral pattern~\cite{ReF:17}. The system evolves looking for a lower pressure zone (new eye position) in the search space. Once a new lower pressure is discovered, its position becomes the eye and the process starts over again~\cite{ReF:17}. 

As the increasing $z_{k}[n]$ on the $k$th wind parcel at the $n$th iteration is unknown, in the traditional HSO~\cite{ReF:17} it is adopted a random variable with uniform distribution, i.e., $z_{k}[n] \sim U \in \mathcal{U} [0, 1]$. However, in this work we propose a  chaotic mechanism combined to HSO (namely hereafter C-HSO) in which a one dimensional {\it logistic map} assign random values to $z_{k}[n]$. Such chaotic logistic map is related to the dynamics of the biological population with the chaotic distribution features \cite{ref:16}-\cite{ReF:103}, obtained by the recursive equation:
\begin{equation}\label{eq:3}
z_{k}[n+1] = \displaystyle \mu \cdot z_{k}[n] (1-z_{k}[n]),
\end{equation}
where ${z_{k}[n]} \in [0, 1]$ is the chaotic variable and $\mu$ is the control parameter in the range $0< \mu \leq 4$~\cite{ref:16} \cite{ReF:103}. The assumed $z_{k}[n]$ values brings randomness to the search step when compared with uniform distribution.

From eq.~\eqref{eq:1} and~\eqref{eq:3}, the power updating of two consecutive channels associated to the $k$th wind parcel at the $n$th iteration is given by:
\begin{equation}\label{eq:2}
\begin{array}{c}
{p}_{k,i}[n]=r_k({\theta_{k}[n]}) \cos({\theta_{k}[n]}) +\hat{p}_{i}[n] \\
{p}_{k,i+1}[n]=r_k({\theta_{k}[n]}) \sin({\theta_{k}[n]}) + \hat{p}_{i+1}[n]
\end{array}
\end{equation}
where $i= (k {\,\,\rm mod\,\,} h) +1$ corresponds to the $i$th user from the $k{\rm th}$ parcel updating, $\rm mod$ is the modulo operator and $h=M-1$ is the number of groups that represents $K$ wind parcels. Each group is denoted by $\mathcal{G}_{i}$, representing the power updating of two specific channels  from ${\bf p}_k$, as in eq.~\eqref{eq:2}, resulting ${\bf p}_k \subset \mathcal{G}_{i}$.

The $\theta_k[n]$ updating from eq.~\eqref{eq:2} is defined by concept of velocity variation of the $k$th wind parcel in the $n$th iteration, which is given by:
\begin{equation}
\label{eq:2a}
\begin{aligned}
\displaystyle \omega_{k}{[n]}=\omega_{\rm max} \cdot \left({\frac{{r}_{k}{[n]}}{p_{\rm max}}}\right) & \quad {\rm if} & \quad {r}_{k}{[n]} < p_{\rm max} 
\\
\displaystyle  \omega_{k}{[n]}=\omega_{\rm max} \cdot \left(\frac{p_{\rm max}}{r_{k}{[n]}}\right)^{z_{k}{[n]}} & \quad {\rm if} & \quad r_{k}{[n]} > p_{\rm max} 
\end{aligned}
\end{equation}
where $\omega_{k}{[n]}$ is a tangential velocity~{of the $k$th wind parcel at $n$th iteration}, $\omega_{\rm max}$ is the maximum tangential velocity {adopted for all the wind parcels}, $p_{\rm max}$ is the transmission power maximum and $z_{k}{[n]}$ is a shape parameter related to the fit data~{at $n$th iteration}~\cite{ReF:17}. Thus, the ${\theta_{k}{[n]}}$ updating at the {$n$th} iteration is given by:  
\begin{equation}
\label{eq:2b}
\begin{aligned}
\displaystyle \theta_{k}{[n]}& =\theta_{k}{[n]}+\omega_{k}{[n]} & \quad {\rm if} & \quad {r}_{k}{[n]} < p_{\rm max} 
\\
\displaystyle \theta_{k}{[n]}& =\theta_{k}{[n]}+\omega_{k}{[n]}\left( \frac{p_{\rm max}}{r_{k}{[n]}}\right)^{z_{k}{[n]}} & \quad {\rm if} & \quad r_{k}{[n]} > p_{\rm max}. 
\end{aligned}
\end{equation}
{As ${p_{\rm max}}$, ${r_{k}[n]}$ and ${z_{k}[n]}$ represent the behavior of updating of $k$th wind parcel, $\omega_{k}$ can be assumed as a fixed value for the $K$ wind parcels, denoted by $\omega$~\cite{ref:16}.}

In addition, the initial power vector of the CHSO is defined as ${\bf p}_0$ while the component ${p}_{{k,i}}$ is defined by the feasible boundaries in the set $ \Omega \in [\,p_{\rm min};\,\,p_{\rm max}\,]$, i.e., the  minimum and maximum transmission (\textsc{tx}) power. Therefore, when ${p}_{{k,i}} \notin  \Omega$, the function $\phi({p}_{{k,i}})$ is true; thus, the initial and current angular coordinates of the  $k$th wind parcel, $\theta_{k}{[1]}$ and $\theta_k$, respectively, must be updated as:
$$
\theta_{k}{[1]}=z_{k}{[1]}\qquad \text{and}\qquad \theta_k=0.
$$
The stopping criterion is defined by the number of iterations $N_f$. A pseudo-code for the CHSO power allocation is described in Algorithm~\ref{algo:1}. 

\begin{algorithm}[!htbp]
\caption{CHSO -- Chaotic Hurricane Search Optimization}\label{algo:1}
\begin{tabular}{ll}
\multicolumn{2}{l}{{{\bf Input:} $N_f$, $K$, $\omega$, $r_0$, $r_{\rm max}$, $\theta_{k}{[1]}$, $p_{\rm min}$, $p_{\rm max}$}, ${\theta_{k}{[n]}}=0$,} \\ 
		\multicolumn{2}{l}{${\bf p}_0;$}    \\ 				
		\multicolumn{2}{l}{{{\bf Output:} $\mathbf{\hat{p}}[n];$}}  \\ 
		{\bf 1:}  & ${\bf \hat{p}}{[n]} = {\bf p}_0$; \\
		{\bf 2:}  & {\bf for} $n=1$ to $N_f$ \\
		{\bf 3:}  & {$p_{\bf \hat{p}}[n]$ = pressure (${\bf \hat{p}}[n]$);} \\				
		{\bf 4:}  &\quad {\bf for} $k=1$ to $K$ \\
		{\bf 5:}  &\quad \quad (a) ${r_k [n]=r_0 \cdot \exp (\theta_{k}{[n]} \cdot z_{k}[n]);}$ \\
		{\bf 6:}  &\quad \quad (b) ${{\bf p}_k [n] = {\bf \hat{p}}[n];}$ \\
		{\bf 7:}  &\quad \quad (c) $\displaystyle {i = (k\,\,{\rm mod}\,\, {h})+1;}$ \\
		{\bf 8:}  &\quad \quad (d) $\displaystyle {{p}_{k,i} [n] =r_i \cdot \cos (\theta_{k}[1]+ {\theta_{k}[n]}) + e_i;}$ \\
		{\bf 9:}  &\quad \quad (e) $\displaystyle {{p}_{k,i+1} [n]= r_i \cdot \sin (\theta_{k}[1]+{\theta_{k}[n]}) + e_{i+1};}$  \\
		{\bf 10:} &\quad \quad (f) {$p_{{\bf P}_k}[n]$ = pressure (${\bf p}_k[n]$);}\\
		{\bf 11:} &\quad \quad (g) {\bf if} $\varphi{({p}_{k,i})}$ {\bf or} $\varphi{({p}_{k,i+1})}$;\\		
		{\bf 12:} &\quad \quad \quad \quad $\theta_{k}[1]={z_{k}[n]}\cdot 2 \pi$;\\
		{\bf 13:} &\quad \quad \quad \quad $\theta_k=0$;\\				
		{\bf 14:} &\quad \quad \,\,\,\,\,\,\, {\bf else if} ${p_{\bf \hat{p}}[n]} <p_{{\bf p}_k}[n]$\\
		{\bf 15:} &\quad \quad \quad \quad ${\bf e}={\bf p}_k$;\\
		{\bf 16:} &\quad \quad \quad \quad $p_{\bf \hat{p}}[n]$ = pressure (${\bf \hat{p}}[n]$);\\		
		{\bf 17:} &\quad \quad \,\,\,\,\,\,\, {\bf else} \\
		{\bf 18:} &\quad \quad \,\,\,\,\,\,\, \quad {\bf if} {$r_{k}[n]<p_{\rm max};$} \\
		{\bf 19:} &\quad \quad \,\,\,\,\,\,\, \qquad {$\theta_{k}[n]=\theta_{k}[n]+\omega;$} \\
		{\bf 20:} &\quad \quad \,\,\,\,\,\,\, \quad {\bf else} \\
		{\bf 21:} &\quad \quad \,\,\,\,\,\,\, \qquad {$\theta_{k}[n]=\theta_{k}[n]+\omega\left( \frac{r_{\rm max}}{r_{k}[n]}\right)^{z_{k}[n]};$} \\
		{\bf 22:} &\quad \quad \,\,\,\,\,\,\, \quad {\bf end} \\
		{\bf 23:} &\quad \,\,\,\,\,\,\, \quad {\bf end} \\
		{\bf 24:} &\quad {\bf end} \\
		{\bf 25:} & {\bf end} \\				
	\end{tabular}
\end{algorithm}

Finally, the quality of the power allocation solution at $n$th iteration (${\bf \hat{p}}[n]$) can be measured by the normalized mean square error (${\rm NMSE}$) related to the optimal solution vector ${\bf p }^*$:
\begin{equation}\label{NMSE}
{\rm NMSE}{[n]}=\mathbb{E}\left[\frac{\| {\bf \hat{p}}{[n]} - {\bf p}^*\|^2 }{\|{\bf p}^*\|^2}\right]
\end{equation}
where $\mathbb{E}$ is the expectation operator and $\| \cdot \|$ is the Euclidean distance to the origin. Herein, the optimal power allocation vector ${\bf p }^*$ is defined by the gradient descent method described in \cite{ref:15}.

\subsection{Complexity Analysis}\label{Complexity}
The computational complexity of the algorithms is calculated following ~\cite{ReF:27a}~\cite{ReF:34}. It is evaluated by amount of execution time as a function of the number of mathematical operations necessary to run until convergence. The number of operation executed includes addition, subtraction, multiplication, division (or mod operator), natural logarithm, power or exponential and trigonometric functions, where each is assumed as one floating-points operation (flop). Logical ({i.e.}, and, or) and comparison ({i.e.}, if, else, else if, $\leq$, etc...) operations, and variable assignment were considered irrelevant time-consuming operations. Hence, the  computational complexity is affected by the number of active channels ($M$), by the size and number of routes, {i.e.}, $N_i^{\textsc{roadm}}$ and $N_i^{\rm span}$, which is related to measured SNR, from eq.~\eqref{eq:9}, as well as by the number of iterations from algorithms $N_f$. Hence, the CHSO and HSO complexity can be defined from Algorithm~\ref{algo:1}, eq.~\eqref{eq:5}, chaotic maps and uniform distribution, resulting:

\begin{equation}\label{C_HSO}
\begin{array}{c}
\mathcal{C}^\text{HSO} = 22 N_f \cdot K + 9 N_f \cdot K + \\
3\cdot \left(19M^2 + 5M + \sum_{i=1}^M(N_i^{\textsc{roadm}}+N_i^{\rm span})\right) N_f \cdot K
\end{array}	
\end{equation}	
and
\begin{equation}
\label{C_CHSO}
\begin{array}{c}
\mathcal{C}^\text{CHSO}
=   22 N_f \cdot K + 3 N_f \cdot K + \\
+  3\cdot \left(19M^2 + 5M+ \sum_{i=1}^M(N_i^{\textsc{roadm}}+N_i^{\rm span})\right) N_f \cdot K,
\end{array}	
\end{equation} 

Asymptotically, the complexity of both algorithms is of order of $\mathcal{O} (M^2)$. Moreover, aiming at  performing a more representative comparison, the complexity of gradient descent (GD) method is also evaluated. It is based on the outline of a general GD method, which defines a descent direction $\Delta_{\bf p}$ and a suitable step size selection using backtracking line search method (from Algorithm 9.1 and  9.2~of \cite{ReF:36}). Here, $\Delta_{\bf p}$ is normalized by $||{J}_1({\bf p})||$. The GD algorithm complexity is given by:
\begin{equation}
\label{C_GD}
\begin{array}{c}
\displaystyle  \mathcal{C}^{\rm GD}
=  N_{f}^{\rm GD} (M^2 + 4M+ 3) + \\
\left[19M^2 + 5M+ \sum_{i=1}^M(N_i^{\textsc{roadm}}+N_i^{\rm span})\right] \\
\cdot \left[N_{f}^{\rm GD}  (5 \cdot N_{bt}^{\rm GD} \cdot M + 5\cdot M  + 1 ) \right] \\
\end{array}	
\end{equation}
where, $N_{f}^{\rm GD}$ is the number of iterations from Algorithm 9.1~\cite{ReF:36}, $N_{bt}^{\rm GD}$ is the number of iterations from the backtracking search. Asymptotically, its complexity is of order of $\mathcal{O} (M^3)$.

\subsection{Input Parameters Optimization for the CHSO}
\label{sectioN_framework_optimization}
The framework for the input parameters optimization (IPO) related to the CHSO performance is similar to the systematic proceeding proposed in~\cite{ReF:30}, in which only the main input parameters that affect dramatically algorithm's performance are optimized, { i.e.}, initial power increasing ($r_0$) and angular velocity ($\omega$). After that, the input parameters directly related to the  algorithm's complexity, { i.e.}, wind parcels $K$ and number of iterations $N_f$ are optimized regarding the performance-complexity tradeoff.

The IPO procedure consists of two steps: a) keep $\omega$ fixed and optimizes $r_0$; b) $r_0$ (from first step) is hold fixed while $\omega$ value is optimized. The optimized input parameter values are found by golden-section search method, which finds the minimum of an objective function by successively narrowing the range of values inside feasible range; in other words, it estimates the maximum and minimum values of the input parameter until the best value of $r_0$ and $\omega$ have been found. Both optimization input parameter procedure adopt the same steps; for this reason Algorithm~\ref{algo:2} details only the $r_0$ optimization.
\begin{algorithm}[!htbp]
	\caption{IPO procedure (CHSO)}\label{algo:2}
	\small
	\begin{tabular}{ll}
		\multicolumn{2}{l}{{{\bf Input:} $N_{\rm lps}$, ${\rm tol}_{r_0}$, ${\rm tol}_{\omega}$, $p_{\rm min}$, $p_{\rm max}$, $\omega_{\rm min}$, $\omega_{\rm max}$, $g_s$, {$N_f$, } }} \\ 
		\multicolumn{2}{l}{{$K$, $\omega$, $r_{\rm max}$, $\theta_{k}{[1]}$, $p_{\rm min}$, $p_{\rm max}$, ${\theta_{k}{[n]}}=0$, ${\bf p}_0$;}} \\ 
		\multicolumn{2}{l}{{{\bf Output:} $\omega$, $r_0$;}}  \\ 
		{\bf 1:} &{\bf for} $n_{{\rm lps}}=1$ to $N_{\rm lps}$ \\
		{\bf 2:} &\quad  ${\textbf{if} \,\,} n_{\rm lps} = 1$ \\
		{\bf 3:} &\quad\quad (a) $r_l =\log(p_{\rm min})$;\\
		{\bf 4:} &\quad\quad (b) $r_u=\log(p_{\rm max})$;\\
		{\bf 5:} &\quad ${\textbf{else} \,\,}$\\
		{\bf 6:} &\quad\quad (a) $I_{n_{\rm lps}}= \frac{{\rm min}(|r_0-r_l|,|r_0-r_u|)}{(0.5g_s^{(n_{\rm lps}-2)})}$\\		
		{\bf 7:} &\quad\quad (b) $r_l=\log(r_0)-I_{n_{\rm lps}}/2$;\\
		{\bf 8:} &\quad\quad (c) $r_u=\log(r_0)+I_{n_{\rm lps}}/2$;\\		
		{\bf 9:} &\quad{\bf end}\\
		{\bf 10:} &\quad keeps $\omega$ fixed;\\		
		{\bf 11:}&\quad{\bf while} $\displaystyle |{r}_{l}-{r}_{u}|<{\rm tol}_{r_0}$ \\				
		{\bf 12:}&\quad\quad  {${\textbf{if} \,\,} \displaystyle  \mathcal{E}_1 <  \mathcal{E}_2$} \\
		{\bf 13:}&\quad\quad\quad (a) $r_u=\hat{r}_{2}$; \\
		{\bf 14:}&\quad\quad\quad (b) ${\hat{r_2}} = r_{u}- g_s(r_{u}-r_{l})$; \\		
		{\bf 15:}&\quad\quad {\textbf{else} {$\mathcal{E}_1 >  \mathcal{E}_2$} \,\,}\\
		{\bf 16:}&\quad\quad\quad (a) $r_{l}={\hat{r_1}}$; \\
		{\bf 17:}&\quad\quad\quad (b) ${\hat{r_1}} = r_{u}+ g_s(r_{u}-r_{l})$; \\		
		{\bf 18:}&\quad\quad{\bf end}\\		
		{\bf 19:}&\quad{\bf end}\\
		{\bf 20:}&$r_0=(r_l+r_u)/2$; \\
		{\bf 21:}&  executes $\omega$ optimization analogous to lines {\bf 2} to {\bf 22}; \\		
		{\bf 22:}&{\bf end}\\
	\end{tabular}
\end{algorithm}
From Algorithm~\ref{algo:2}, analogous the golden section search algorithm~\cite{ReF:39}, the golden-section value is $g_s=\frac{1+\sqrt{5}}{2}$, while $r_{l}$ and $r_{u}$ are lower and upper bound of $r_0$, respectively; $N_{\rm lps}$ is the number of loops for reduction of the interval $I_{n_{\rm lps}}$,  ${\hat{r}_1}$ and ${\hat{r}_2}$ are the intermediates points; $|{r}_{l}-{r}_{u}|<{\rm tol}_{r_0}$ is the stopping criterion of $r_0$; ${\rm tol}_{r_0}$ is the tolerance adopted; $\omega \in [\omega_{\rm min}; \omega_{\rm max}]$ is the parameter keeps fixed; and $\log$ operator performs the normalization of $r_0$ range. $\mathcal{E}_1$ and $\mathcal{E}_2$ are given by the $\mathbb{E}$ [${J_1}({\bf \hat{p}}[n])]$, assuming $r_0=10^{(\hat{r}_{1})}$ and $r_0=10^{(\hat{r}_{2})}$, respectively, while ${\bf \hat{p}}[n]$ is calculated via Algorithm 1. $N_r$ realizations are adopted to measure $\mathbb{E}$ [${J_1}({\bf \hat{p}})[n]$]. In this context, the $\omega$ optimization is obtained by replacing the variable $r_0$ by $\omega$ and vice versa.

\section{Numerical Results}\label{Numerical_Results}
In this section, the performance of the CHSO and HSO are analyzed and systematically compared. Section \ref{Section_NetPar} presents network's scenario and parameters, while section ~\ref{Section_IPO} describes the input parameters optimization (IPO). Sections ~\ref{Section_PE_PA_PCC} and~\ref{Section_PE_PA_nPCC} analyse  the power allocation performance for boths CHSO and conventional HSO methods considering perfect and non-perfect channel estimation, respectively. Computational complexity assuming different system loading is discussed in section~\ref{Section_CC}. The numerical simulations were performed with MATLAB (version 7.1) in a computer with 32 GB of RAM and processor Intel Xeon E5-1650 (3.5 GHz).

\subsection{Network Parameters}\label{Section_NetPar}
\begin{figure}[htbp!]
	\centering
	\includegraphics[trim={0mm 0mm 0mm 0mm},clip, width=.75\linewidth]{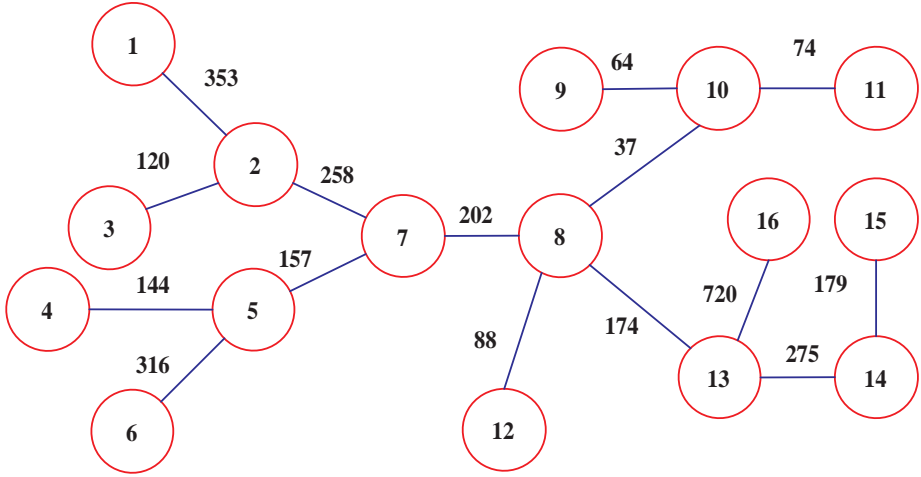}
	\caption{Adopted EON topology; distance in km.}
	\label{fig:3}
\end{figure}

Fig.~\ref{fig:3} illustrates a virtual network topology for the transmission routes from source ($\mathcal{S}$) to destination ($\mathcal{D}$). The span length is 100 Km, channel spacing ($\Delta f$) of 50 GHz and guard band of 6 GHz. This topology was chosen to concentrate the routes ${\mathcal{R}}$ in some links, thus the effects of interference, as well as the effects of nonlinearities are more prominent. The EON transmission capability is in range of 100 to 300 Gbps. The routes and spectrum assignment procedure is out of the scope of this work, as these are considered to be stablished by a routing and spectrum assignment (RSA) algorithm. Bit rate requirement, routes, distance and modulation format are listed in Table \ref{tab:3}. The physical layer parameters values of the elastic optical network are illustrated in Table ~\ref{tab:4},~\cite{ref:2}~\cite{ReF:20}~\cite{ReF:20a}~\cite{ReF:38}~\cite{ReF:35}~\cite{ReF:37}. Herein, we evaluate only twelve channels, including the channels with higher and lower power transmitted power, to avoid burden information.

\begin{table}[htbp!]
	\centering
	\caption{Channels features: Routes, Distance, Bit Rate and Modulation Format.}
	\label{tab:3}
	\scalebox{1}{
		\begin{tabular}{ c c c c c c}
			\hline \hline
			{\bf Route} & {\bf $\mathcal{S} \rightarrow \mathcal{D}$} & {\bf Distance} (km) & ${\bf R}$ (Gbps) & {\bf Modulation}
			\\ \hline \hline
			${\mathcal{R}}_1$       & 1 - 16 &  1707 &  100 & PM-QPSK \\
			${\mathcal{R}}_2$       & 1 - 15 &	1441 &	100 & PM-QPSK \\\hline
			${\mathcal{R}}_3$       & 1 - 14 &	1262 &	100 & PM-QPSK  \\
			${\mathcal{R}}_4$       & 1 - 9  &  914  &  100 & PM-QPSK \\ \hline				
			${\mathcal{R}}_5$       & 3 - 14 &	1029 &	150 & PM-8QAM \\
			${\mathcal{R}}_6$       & 3 - 13 &	754  &	150 & PM-8QAM \\ \hline
			${\mathcal{R}}_7$	    & 3 - 12 &	842  &	200 & PM-16QAM \\
			${\mathcal{R}}_8$       & 6 - 10 &  712  &  200 & PM-16QAM \\ \hline
			${\mathcal{R}}_9$       & 4 - 9  &	604  &	250 & PM-32QAM \\
			${\mathcal{R}}_{10}$	& 5 - 11 &	470  &	250 & PM-32QAM  \\ \hline
			${\mathcal{R}}_{11}$	& 7 - 11 &	235  &	300 & PM-64QAM \\
			${\mathcal{R}}_{12}$	& 7 - 10 &	313  &	300 & PM-64QAM \\			
			\hline \hline
		\end{tabular}
	}	
\end{table}

\begin{table}[htbp!]
	\centering
	\caption{Physical layer parameters}\label{tab:4}
	\scalebox{1}
	{
		\begin{tabular}{lccc}
			\hline
			{\bf Description} & {\bf Variable} & \multicolumn{2}{c}{\bf  Value}\\\hline
			{Bit-error-rate acceptable at pre-FEC}~\cite{ReF:20} & ${\rm BER}^*$ & \multicolumn{2}{c}{$4\cdot 10^{-3}$}\\
			Minimum Tx power  & $p_{\min}$ (dBm) & \multicolumn{2}{c}{ $-100$}\\
			Maximum Tx power  & $p_{\max}$ (dBm) & \multicolumn{2}{c}{$20$ }\\
			Channel spacing   & $\Delta f$ (GHz) & \multicolumn{2}{c}{50} \\
			Planck constant~\cite{ReF:37}   & $h$ (J/Hz) & \multicolumn{2}{c}{$6.6261\cdot 10^{-34}$ }\\ 
			Light frequency~\cite{ReF:37}   & $v_c$ (Hz) & \multicolumn{2}{c}{$193.55 \cdot 10^{12}$ }\\{Group Velocity Dispersion (GVD)}~\cite{ReF:37} & $\beta_2$ (${\rm s}^2/{\rm km}$) & \multicolumn{2}{c}{$2.07\cdot 10^{-23}$} \\
			Nonlinear parameter of the fiber~\cite{ReF:20} & $\gamma$ $({\rm W}/{\rm km})$ & \multicolumn{2}{c}{ $1.3$} \\
			Span length with standard single mode~\cite{ReF:20}& $L$ (km) & \multicolumn{2}{c}{ $100$ } \\
			Uncertain SNR monitoring, 
			~\cite{ReF:38} & $\epsilon_i$ (dB) & \multicolumn{2}{c}{$\sim \mathcal{LN} \displaystyle (\mu, \sigma)$}  \\			{Standard deviation of $\epsilon$}~\cite{ReF:38} & $\sigma$ (dB) & \multicolumn{2}{c}{[0; 0.16]} \\
			{Expectation of  $\epsilon$}~\cite{ReF:38}  & $\mu$ (dB) & \multicolumn{2}{c}{0} \\				
			{Margin Residual tolerance for lower bound}  & \multirow{2}{*}{$\varLambda_1$} &  \multicolumn{2}{c}{\multirow{2}{*}{$4$E$-$3} } \\
			{Tolerance adopted for the upper bound}  & \multirow{2}{*}{$\varLambda_2$} &  \multicolumn{2}{c}{\multirow{2}{*}{$1$E$-$3} } \\
			{of the residual margin}\\					{Maximum power perturbation}~\cite{ReF:35} & {$A_{\rm pert}$ (dB)} & \multicolumn{2}{c}{1} \\
			{Begin-of-Life (BoL)} & $\tau_0$ (years) & \multicolumn{2}{c}{0} \\
			{End-of-Life (EoL)} & $\tau_{\rm end}$  (years) & \multicolumn{2}{c}{10} \\
			\hline
			{\bf Equipment Ageing Effect} & & {\bf BoL} & {\bf EoL} \\ \hline
			{Fiber loss coefficient}~\cite{ref:2} & $\alpha_f$ (dB/km) & $ 0.22$ & $0.23$ \\ 
			{Connector Loss}~\cite{ref:2} & $c_{\rm loss}$ (dB) & $0.20$ & $0.30$ \\ 
			{Connectors per span}~\cite{ref:2} & $s_{\rm loss}$ & 2 & 2 \\ 				
			{Splice Loss}~\cite{ref:2} & $s_{\rm loss}$ (dB) & $ 0.30$ & $0.50$ \\ 
			{Number of splices}~\cite{ref:2}  & $s_e$ (km$^{-1}$) & $2$ & $2$ \\ 		{EDFA noise figure}~\cite{ref:2} & $N_e$ (dB) & $4.50$ & $5.50$ \\ 	
			ROADM loss 
			~\cite{ref:2} & $A^{\rm ROADM}$ (dB) & $20.0$ & $23.0$ \\
			{Transponder Margin}~\cite{ref:2} & $M_t$ (dB) & $ 1.00$ & $1.50$	\\
			{Design Margin}~\cite{ref:2} & $M_d$ (dB) & $2.00$ & $1.00$\\ 
			\hline
		\end{tabular}
	}
\end{table}

\subsection{IPO Procedure under Perfect Channel Conditions}\label{Section_IPO}
This step is very important for EON operation under all the operation conditions, such as uncertainty of SNR monitoring, effects of ageing and power instability. For this reason, the IPO-performance and IPO-Complexity are treated in the next subsections (\ref{PerformanceIPO} to~\ref{trade_offIPO}),
assuming the  EON operating under perfect channel conditions, which is given by: perfect estimation of SNR, operation at the BoL and static scenario, following the Table~\ref{tab:3} and~\ref{tab:4}, for operation at any conditions. The round-trip delay are compensated from traditional Smith predictor~\cite{ReF:34}.

Basically, there are four main input parameters, which can be divided into two groups: input parameters that affect directly the performance, given by initial value of the power increasing $r_0$ and the tangential velocity for the power increasing $\omega$; and input parameters that affect directly the HSO algorithm complexity, given by wind parcels $K$ and iterations number $N_f$. The optimization of both groups is discussed in the subsections~\ref{PerformanceIPO} and~\ref{ComplexityIPO}, respectively. Others input parameters are described in Table~\ref{tab:5}. Finally, the IPO under the perspective of complexity-performance tradeoff is elaborated in subsection~\ref{trade_offIPO}.
{\begin{table}[htbp!] 
		\centering
		\caption{CHSO and HSO parameters}\label{tab:5}
		\scalebox{.95}
		{\begin{tabular}{lll}
				\hline
				{\bf Param.} & {\bf Description} & {\bf Value} \\ \hline
				{$\omega_{\rm min}$} & Minimum angular velocity & $10^{-4} \cdot \pi$\\
				{$\omega_{\rm max}$}  & Maximum angular velocity & 2 $\pi$ \\
				{$\omega$} & Angular velocity & $[10^{-3}; 2\pi]$ \\
				{$r_0$} & power increasing [dBm] & $[-100; 20]$  \\
				$M$ & Search space dimension or channels number & $12$\\
				$K$ & Wind parcels number & $M \cdot N_w  $
				\\	
				$N_{f}$ & Number of iterations & $[100; 500]$
				\\
				$N_{\rm lps}$ & number of loops in the IPO procedure& 30
				\\
				$N_r$ & Number of realizations & 100
				\\
				$N_w$ & Wind parcels factor & $[1; 20]$				
				\\			
				$\theta_{k,1}$ & Initial angles of the $k$th wind-parcel & 0
				\\
				$\theta_{k}$ & Angles of the $k$th wind-parcel & $z_{k,n}$				
				\\
				${\bf \hat{p}}$ & Initial eye (dBm)  & 0
				\\ 
				$z_{k,n}$ & Chaotic variable & $[0; 1]$\\
				$\mu$ & Control variable of the chaotic logistic map & $4$\\
				$\upsilon$ &  ${J}_1$ weighting  & 1  \\				
				\hline
			\end{tabular}
		}
\end{table}}	

\subsubsection{IPO-performance under Perfect Channel Conditions}\label{PerformanceIPO}
{In this context, $r_0$ and $\omega$ affect drastically the algorithm's performance. The optimized values are obtained by the framework previously described in section~\ref{sectioN_framework_optimization}. It assumed $\omega=1.5708$ as an initial value for the tangential velocity, {number of wind parcels $K=180$} and number of iterations equal to $N_f=250$, all defined empirically.}

\begin{figure}[htbp!]
	\centering
	\includegraphics[trim={10mm 5mm 10mm 5mm},clip, width=1\linewidth]{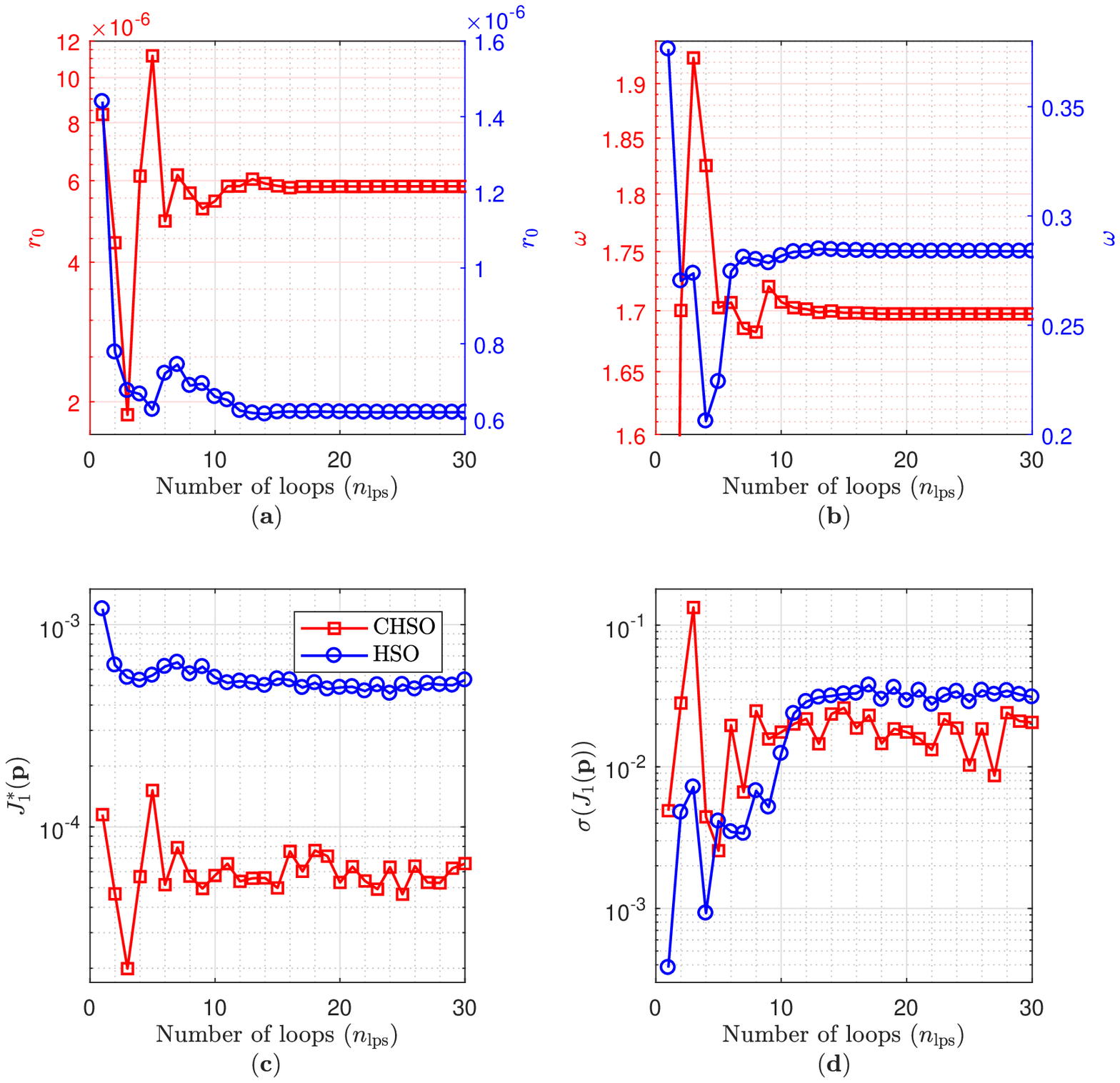}
	\caption{Input parameters optimization for the HSO and CHSO. {\bf a}) $\omega$ optimization; {\bf b}) $r_0$ optimization; {\bf c}) the best value of cost function ${J_1^*}({\bf p})$ in $N_r$ realizations;  {\bf d}) standard deviation for $J_1({\bf p})$.}
	\label{fig:4}
	\vspace{-4mm}
\end{figure}

Fig.~\ref{fig:4} illustrates the $r_0$ and $\omega$ optimization across the loops, in such a way that all the optimized parameters reach full convergence. Different $\omega$ and $r_0$ values were obtained for both algorithms in Fig.~\ref{fig:4}.a) and  ~\ref{fig:4}.b), demonstrating that the higher parameters values from CHSO perform more accelerated and exploitive (via map chaotic) searches. Consequently, the CHSO found a better solutions, measured by the cost function  $J_1({\bf p})$ during $N_r$ realizations and their respective standard deviation, as depicted in  Figs.~\ref{fig:4}.c) and~\ref{fig:4}.d). More details are listed in Table~\ref{tab:6}, considering three loops that describe the optimization trend, { i.e.}, $n_{\rm lps}\in [1; \,15; \, 30]$; the finals optimized parameters is  highlighted by bold face, while the parameters kept fixed at each loop is underlined.

{\begin{table}[htbp!]
		\centering
		\caption{Performance of the IPO procedure for the CHSO and HSO\label{tab:6}}
		\scalebox{0.97}
		{
			\begin{tabular}{cccccc}
				\hline
				$n_{{\rm lps}}$ 	& Alg. 		& $r_0$ 			& $\omega$ 			& ${J_1^*}({\bf p})$ 	& $\displaystyle \sigma{(J_1({\bf p})| \omega,r_0)}$ \\ \hline
				\multirow{2}{*}{1} 	& CHSO 		& 8.3311E-06 		& {\ul {1.5708}}	& 1.1493E-04 			& 4.8946E-03 	\\
				& HSO 		& 1.4398E-06  		& {\ul {1.5708}} 	& 1.1961E-03 			& 3.8201E-04 	\\
				\multirow{2}{*}{1}	& CHSO 		& {\ul {8.3311E-06}}& {9.1054E-01}  	& 1.1493E-04 			& 4.8946E-03 	\\
				& HSO 		& {\ul{1.4398E-06}} & 3.7636E-01     	& 1.1961E-03    		& 3.8201E-04    \\ \hline
				- 					& $\vdots$ 	& $\vdots$ 			& $\vdots$ 			& $\vdots$ 				& $\vdots$		\\ \hline
				\multirow{2}{*}{15}	& CHSO 		& 5.8452E-06        & {\ul{1.6998}}  	& 4.9901E-05    		& 2.6051E-02    \\
				& HSO 		& 6.1944E-07        & {\ul{2.8458E-01}} & 5.3867E-04			& 3.2671E-02    \\
				\multirow{2}{*}{15} & CHSO 		& {\ul {5.8452E-06}}& 1.6982	        & 4.9901E-05   			& 2.6051E-02    \\
				& HSO 		& {\ul {6.1944E-07}}& 2.8418E-01        & 5.3867E-04    		& 3.2671E-02    \\ \hline	
				- 					& $\vdots$ 	& $\vdots$ 			& $\vdots$ 			& $\vdots$ 				& $\vdots$ 		\\ \hline	
				\multirow{2}{*}{30} & CHSO 		& 5.8318E-06        & {\ul{1.6975}}   	& 6.2371E-05   			& 2.1000E-02    \\
				& HSO 		& 6.1873E-07        & {\ul{2.8386E-01}}	& 5.0139E-04   			& 3.2378E-02    \\
				\multirow{2}{*}{30} & CHSO 		& {\ul{\bf 5.8318E-06}} & {\bf 1.6975}        	& {6.5770E-05}    		& {2.0588E-02}    \\
				& HSO 		& {\ul{\bf 6.1873E-07}} & {\bf 2.8386E-01}      & {5.3229E-04 }   		& {3.1103E-02}   	\\
				\hline         		
			\end{tabular}
		}
\end{table}} 

In addition to the proposed optimization by the framework, we perform a numerical analysis of the conditional probability of success (CPoS), which is the probability of $M$-users to achieve the ${\rm BER}^*$ in the direction of the lower budget power given $n$ and $r_0$, denoted by $\mathcal{P}s_1$. In this numerical analysis, $n \in [1; N_f]$ and $r_0 \in [10^{-8}; 10^{-4}]$, both from Table~\ref{tab:6}. The $\omega$ parameter is not evaluated, so it assumes the optimized value from the Table~\ref{tab:6}. 

The formulation for $\mathcal{P}s_1$ follows the RM concept discussed in eq.~\eqref{eq:5}. Then, given $r_0$ and $n$, $\mathcal{P}s_1$ can be defined as the probability of $M$-users to satisfy two conditions: {\bf i}) $\Psi_i \geq \Psi^*-\varLambda_1$, where $\varLambda_1$ assures the ${\rm BER}^*$; 
{\bf ii}) $\Psi_i  \leq \Psi^*+\varLambda_2$, where $\varLambda_2$ assures the ${\rm BER}^*$,  
implying in a  ${\Psi}_i=10\log10(1.001)$  = 4.341$\cdot10^{-3}$ dB for the EON system of Table \ref{tab:6}. In this context, $\mathcal{P}s_1$ is given by:
\begin{equation}\label{eq:14}
\mathcal{P}s_1  \buildrel \Delta \over = {\rm Pr} [{\rm \Psi^*}-{\rm \varLambda}_1 \leq {\rm \Psi} \leq \Psi^*+ {\rm \varLambda}_2 \,\, | r_0,n].
\end{equation}

\begin{figure}[htbp!]
	\vspace{-1mm}
	\centering
	\includegraphics[trim={0mm 0mm 0mm 0mm},clip, width=1\columnwidth]{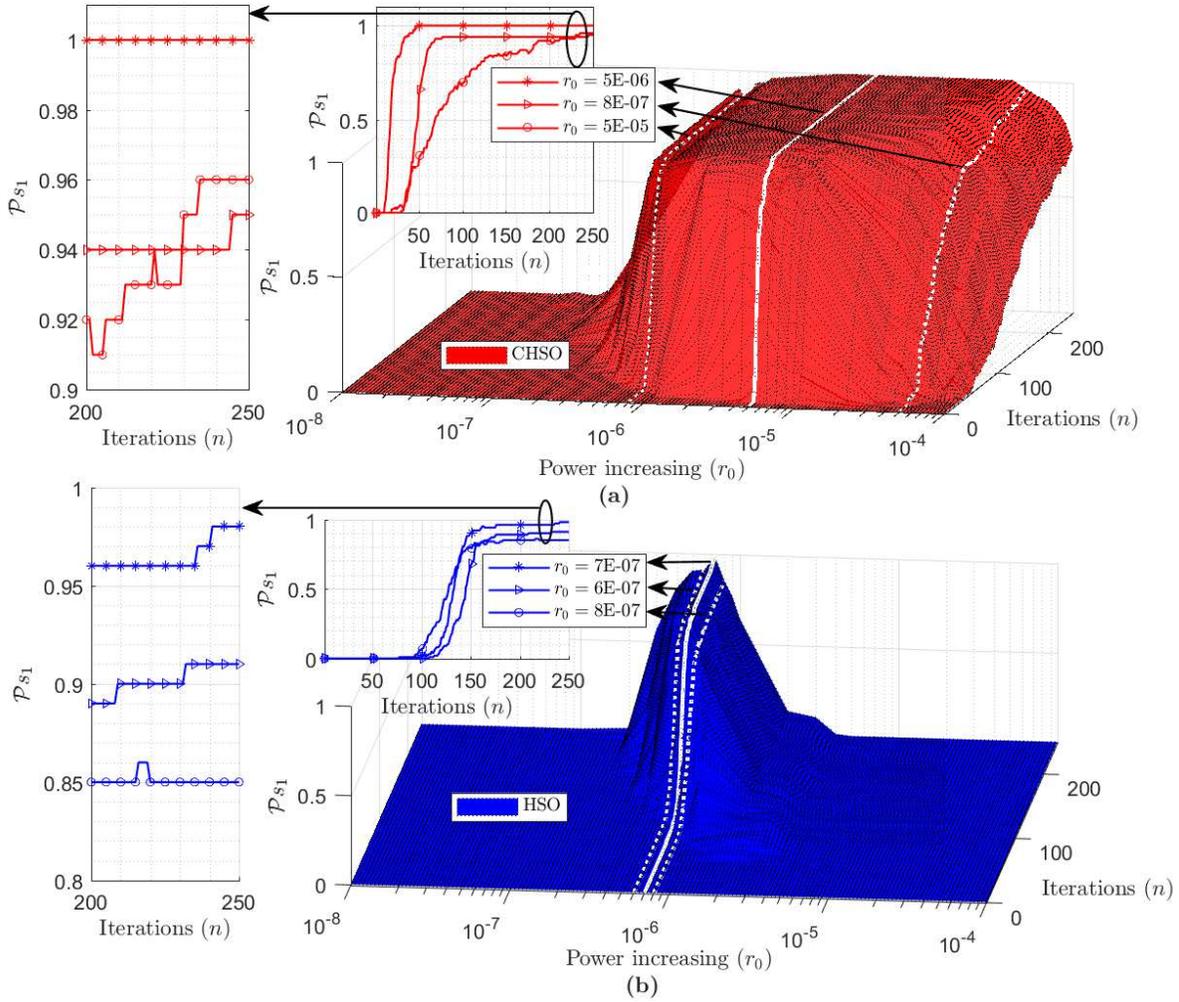}
	\vspace{-7mm}
	\caption{Conditional probability of success ($\mathcal{P}s_1$):  {\bf a}) CHSO; {\bf b}) HSO.}
	\label{fig:6}
\end{figure}

Fig.~\ref{fig:6} depicts the conditional probability of success $\mathcal{P}s_1$ as a function of $r_0$ and number of iterations  from the CHSO and HSO, assuming an average behavior over $N_r$ realizations. Both strategies have attained success, defined as $\mathcal{P}s_1 \geq 0.94$. In the case of CHSO,  a wider range of success regarding HSO has been achieved, defined by $r_0 \in [5\cdot 10^{-6};\,\,5\cdot 10^{-5}]$, and showing that the algorithm presents robustness and lower sensibility during the IPO procedure. The best value for the CHSO input parameter  is obtained as $r_0^*=5\cdot 10^{-6}$, achieving fast convergence ($n=50$) and superior performance, {\it i.e.,} $\mathcal{P}s_1=1$. On the other hand, under HSO, the CPoS is found for a narrow range of power increment, $r_0^* \in [6 \pm 0.5] \cdot 10^{-7} $, because adopting similar values, such as $r_0=6\cdot 10^{-7}$ or $r_0=8\cdot 10^{-7}$ did not allow HSO  achieve $\mathcal{P}s_1 \geq 0.94$. Hence, HSO presented lower robustness and greater sensibility in adjusting its input parameter in the IPO step. Besides, the HSO found slower convergence and worse performance: $\mathcal{P}s_1(50)=0$; and $\mathcal{P}s_1(250)=0.98$, both at $r_0^*$. 

Summarizing, the best $r_0$ and $\omega$ parameters found are registered in the last row of Table~\ref{tab:6}. Varying $r_0$ with fixed $\omega$, $\mathcal{P}s_1$ found a range of $r_0$ that achieved success for both algorithms. This range defines the ability of updating power, which is directly related to robustness from both algorithms. Hereafter, we adopt for any condition of network's operation: $r_0 \in [5\cdot10^{-6};\,\,5 \cdot 10^{-5}]$ and $\omega=1.6975$ for the CHSO; and $r_0 \in [6 \pm 0.5]\cdot 10^{-7}$ and  $\omega=2.8386\cdot 10^{-1}$ for HSO. 
\subsubsection{IPO-complexity under Perfect Channel Conditions}\label{ComplexityIPO}	
$K$ and $N_f$ are the parameters that affect drastically the algorithm's complexity. Thus, analogous to $\mathcal{P}s_1$, the optimization of theses parameters is modelled by:
\begin{equation}
\label{eq:15}
\displaystyle {\mathcal{P}s_2  \buildrel \Delta \over = {\rm Pr} [{\rm \Psi^*}-{\rm \varLambda}_1 \leq {\rm \Psi} \leq \Psi^*+ {\rm \varLambda}_2 \,\,  | K,N_f,r_0,\omega],}
\end{equation}
where from previous subsection, it was adopted $r_0=5.8318 \cdot 10^{-6}$ and $\omega=1.6975$, for the CHSO; and $r_0= 6.1873\cdot 10^{-7}$ and $\omega=2.8386\cdot 10^{-1}$, for the HSO. 

\begin{figure}[htbp!]
	\vspace{-1mm}
	\centering
	\includegraphics[trim={1mm 0mm 4mm 0mm},clip, width=1\columnwidth]{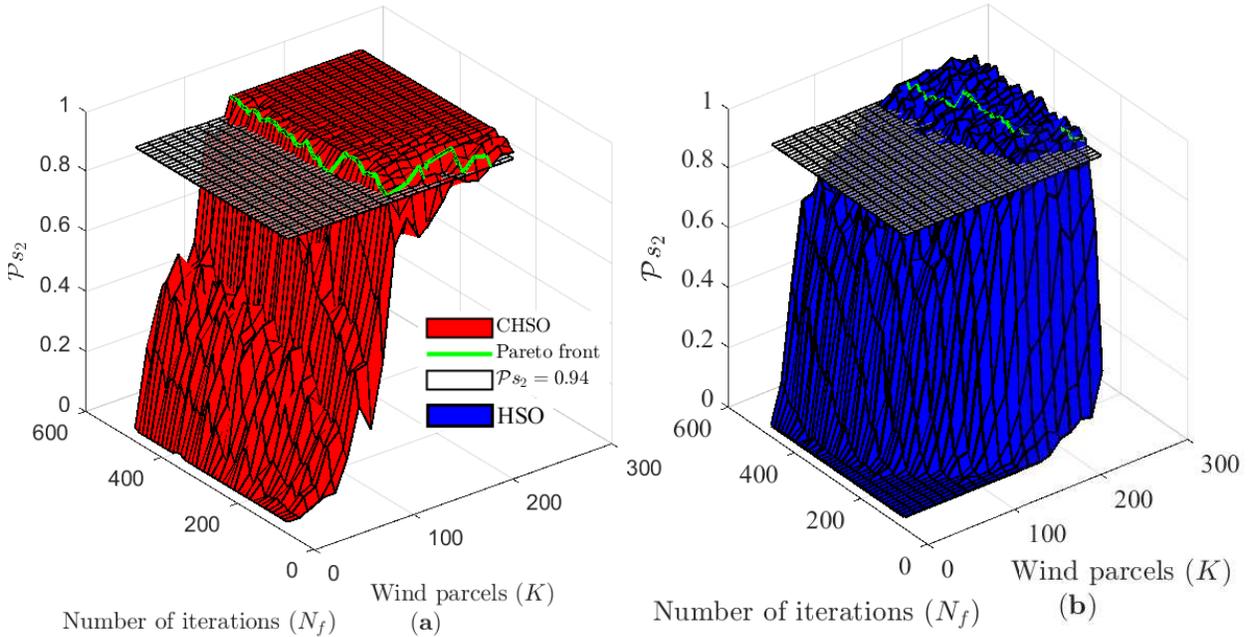}
	\vspace{-7mm}
	\caption{Conditional probability of success  $\mathcal{P}s_2$:  {\bf a}) Top View CHSO ({\bf a}) CHSO and ({\bf b}) HSO.}
	\label{fig:10}
\end{figure}

Fig~\ref{fig:10} depicts  $\mathcal{P}s_2$ from both algorithms, assuming an average behavior of $N_r$ realizations. As can be observed, a set of infinite number of pairs  $(K;\,N_f)$ values combinations found the CPoS, defined as $\mathcal{P}s_2\geq0.94$. Thus, to highlight the reliable and feasible region, Fig~\ref{fig:10}. a)~and~\ref{fig:10}. b)  illustrate (green curve) the Pareto {frontier} (PF). The PF is composed by all success points $(K^*;\,N_f^*)$ assumed as reliable and viable. Hence, all the success points $(K;\,N_f)$ is defined by the set
$$
\mathcal{V} = \{K\in \mathcal{K},\,\, \text{and}  \,\,\, N_f \in \mathcal{N}_f\}\,\,\, |\,\,\, \mathcal{P}s_2 \geq 0.94,
$$ 
while the PF subset $\{(K_\iota^*, {N_f}^*_\iota)\}$ can be defined as:
\begin{eqnarray}\label{eq_PF_1}
& \forall \, (K;\, {N_f}) \in \mathcal{V}  \,\,\, | \,\,\, \forall \,{N_f}_i, \notag\\
\hspace{-8mm}&  \hspace{-8mm} K_\iota^* = \min ({N_f}_i \cdot K_j \, | \,K_j\geq K^*_{\iota-1})    \,\, \text{and}\,\,\, {N_f}^*_\iota={N_f}_i 
\end{eqnarray}
where all $(K^*_\iota;\, {N_f}_\iota^*)$ result of the increasing of $i=[1,\cdots,N_{N_f}]$ and $j=[1,\cdots,N_{K}]$, that represent the decreasing of $K$ and $N_f$, respectively, with $N_{N_f} = | \mathcal{N}_f|$ and $N_{K}=|\mathcal{K}|$.

In terms of PF, the CHSO results are better than HSO, showing a wider region for valid pairs $(N_f; K)$, while providing higher regularity in the plane that corresponds to the reliable and feasible region, combined to lower pairs values. 

\subsubsection{Performance-Complexity Tradeoff}\label{trade_offIPO}
Under channel perfect conditions, the group of input parameters $\omega$, $r_0$, $K$ and $N_f$ should be defined in terms of performance-complexity tradeoff; mathematically it can be modelled as:
\begin{equation}
\mathop{{\text{min}}}\limits_{K,N_f} \left(\mathcal{C}(N_f,K)| \mathcal{P}s_2  \geq 0.94,r_0,\omega \right)
\end{equation}
where $\mathcal{C}(\cdot)$ is the computational complexity for the CHSO or HSO. The feasible solutions are given by the optimized values of $r_0$ and $\omega$, and Pareto front obtained from the pairs ($K$, $N_f$) in Fig~\ref{fig:10}. As a result, we have found a better performance-complexity tradeoff for the CHSO regarding the HSO, where the best solution for the CHSO is defined as $K=132$ and ${N_f}=180$, {i.e.},  $\mathcal{C}^\text{CHSO}=17.3705$ M flops. While the best solution for the HSO is defined by $K=228$ and $N_f=150$,  { i.e.}, $\mathcal{C}^\text{{HSO}}=24.986$ M flops. This IPO framework is summarized in Table~\ref{tab:9}.

\begin{table}[htbp!]
	\caption{CHSO and HSO - Optimized Input Parameters and respective complexity \label{tab:9}}
	\centering
	\scalebox{.95}
	{\begin{tabular}{c|cccc|c}
			\hline
			\bf Algorithm & $r_0$ &$\omega$ & $K$ & $N_f$ & $\mathcal{C}$ [Mflops] \\
			\hline\hline
			CHSO & $[5.10^{-6};\,\,5.10^{-5}]$ & 1.6975		& 132       & 180   &     $17.371$       \\
			HSO  & $[6 \pm 0.5]\cdot 10^{-7}$ & 0.2839 	& 228       & 150   &  $24.986$     \\
			\hline
		\end{tabular}
	}
\end{table}	

\subsection{Power Allocation under Perfect Channel Conditions}
\label{Section_PE_PA_PCC}
\begin{figure}[htbp!]
	\centering
	\includegraphics[trim={5mm 0mm 8mm 5mm},clip, width=1\linewidth]{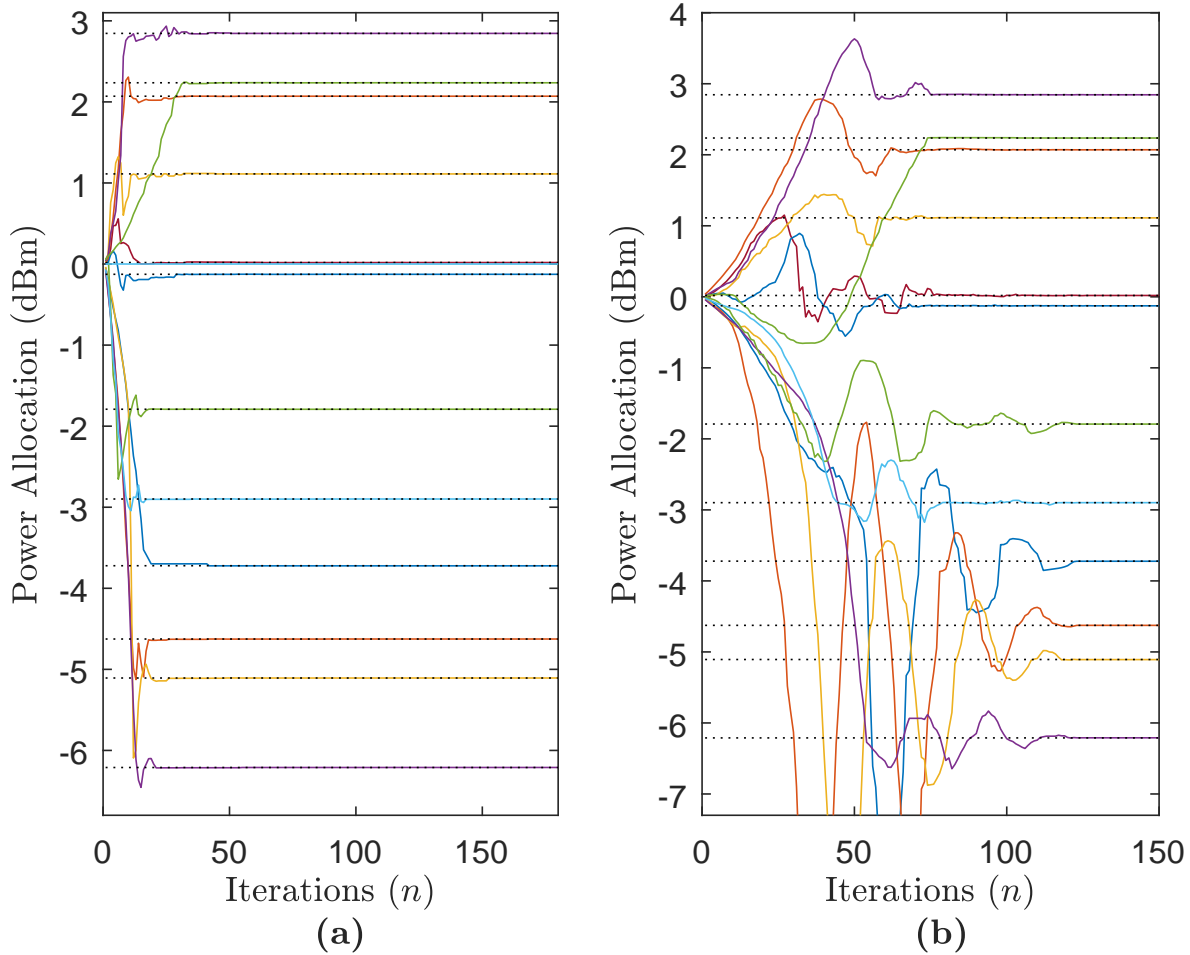}
	\caption{Power allocation per channel {\it versus} the number of iterations: {\bf a}) CHSO; {\bf b}) HSO. Dashed lines in both graphs represent GD solution.}
	\label{fig:7}
\end{figure}

Assuming IPO procedure has been performed previously, the power allocation per channel across iterations can be obtained, as illustrated in Fig.~\ref{fig:7}. In the simulations, it has been assumed perfect channel estimation, optical network operating at the BoL and static scenario, with routes, distances and bit rates given in Table~\ref{tab:3}, as well as physical parameters values following Table~\ref{tab:4}. The general parameters of the algorithms are adopted from the Table~\ref{tab:5}, while performance and complexity parameters are adopted from the Table~\ref{tab:9}, being $r_0=5.8318\cdot 10^{-6}$ (CHSO) and $r_0=6.1873\cdot 10^{-7}$ (HSO). Indeed, the power allocation per channel reaches full convergence for both algorithms. The horizontal dashed lines represent the power allocation per channel obtained via gradient descent procedure, which is an analytical method that has been used to validate convergence of both hurricane heuristic methods.

Regarding the results in Fig.~\ref{fig:7}, the following metrics have been calculated to the CHSO and HSO: {\it a}) the mean integral absolute value of the residual margin for the $M$-channels during time-window resulted equal to 19.1287 dB and 23.1334 dB, respectively; the maximum PP all the channels (${\rm max} (\overline{\bf p})$) at the last iteration of {3.3811$\cdot 10^{-4}$} dB and {1.4014$\cdot 10^{-3}$} dB, respectively; {\it b}) mean settling iteration of all the users ($\overline{i_s}$), assuming tolerance around $10^{-4}$ for the $M$ channels ({i.e.}, ${\bf p^*}-{\bf p}\leq${$1\cdot 10^{-7}$}), results in $\approx$ 79 and $\approx$ 129 iterations, respectively. In this sense, the superiority from the CHSO is evident. Besides, Fig.~\ref{fig:7} presents overshooting and undershooting during the power allocation, which is much more noticiable  in the HSO convergence. This behaviour is called sub-damped, where the transient responses are oscillatory and the closed-loop poles are complex conjugates. 

\begin{figure}[htbp!]
	\centering
	\includegraphics[trim={1mm 7mm 2mm 0mm},clip, width=.8\linewidth]{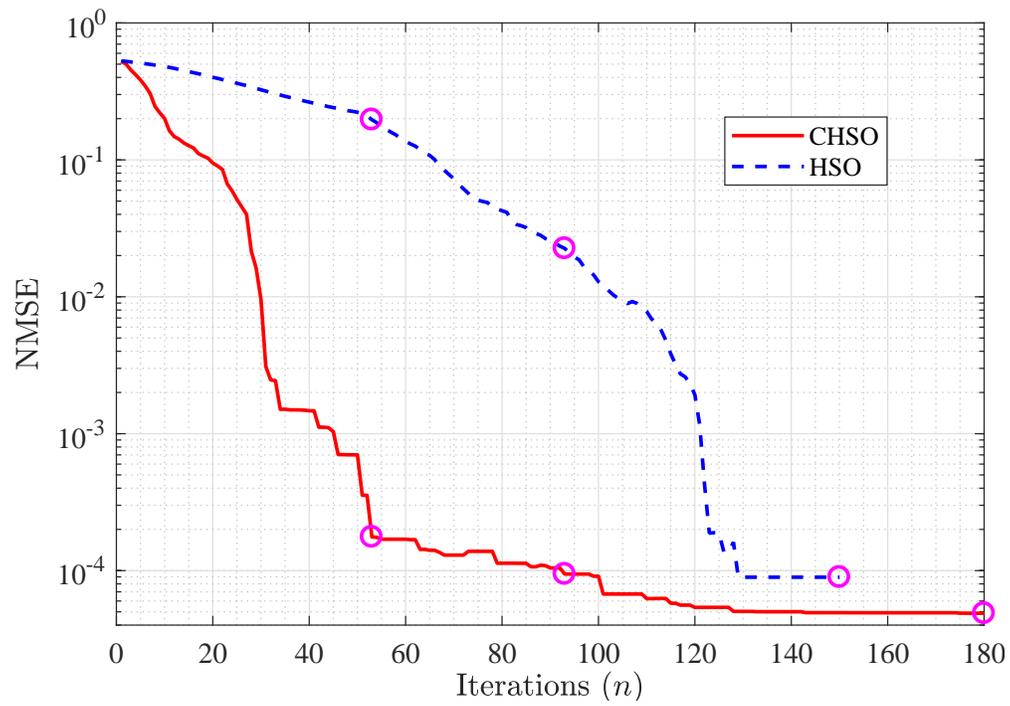}\\
	({\bf a}) Perfect channel conditions\\
	\includegraphics[trim={1mm 7mm 2mm 0mm},clip, width=.80\linewidth]{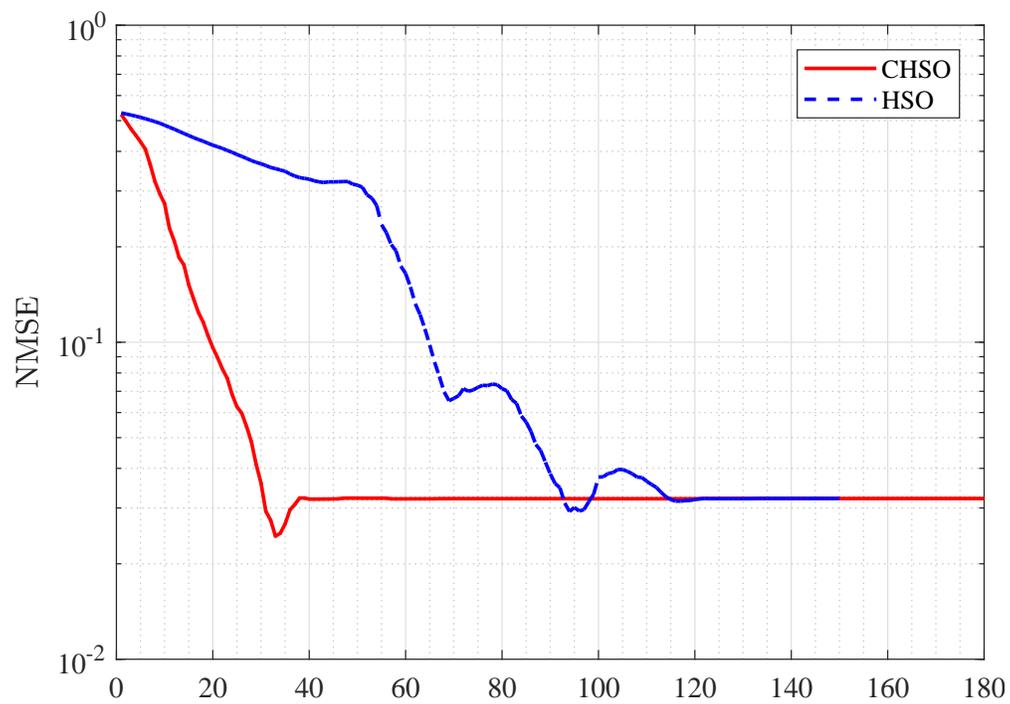}\\
	({\bf b}) Imperfect channel conditions
	\caption{Normalized mean square error ($\textsc{nmse}$) against the number of iterations for CHSO and HSO algorithm operating under perfect and imperfect channel conditions.}
	\label{fig:8}
\end{figure}

Fig.~\ref{fig:8}.a) depicts the quality of the solution by the ${\rm NMSE}$ analysis from Fig.~\ref{fig:7}. In this figure, three main behaviors are highlighted through the circles $c_1$, $c_2$ and $c_3$. The point $c_1 (n=53)$ represents the ability of the CHSO to find a better candidate solution in few iterations, {\it i.e.}, it found a NMSE $= 1.76 \cdot 10^{-4}$, while the HSO was able to attain NMSE$=  0.1965$.  The intermediate point $c_2$ ($n=93$) represents the CHSO around a good candidate solution, in consequence of its more exploitative nature, it is a region which the CHSO can be slower than HSO. In this region, the NMSE reduction are of order of 2.0451$\cdot 10^{-6}$ and 9.4385$\cdot 10^{-5}$, for CHSO; and 4.3480$\cdot 10^{-3}$ and $=2.2552$$\cdot 10^{-2}$, for HSO. $c_3$ represents the CHSO ability to achieve a better solution at the last iteration: the CHSO found a NMSE$=4.87768$$\cdot 10^{-5}$; and HSO found a NMSE$=8.9501$$\cdot 10^{-5}$ for the HSO. Besides, in range of $n=1$ to 53 is evident the instability by HSO, due to its lower exploitative capacity for the power launch (or initial power of the eye) of 0 dbm. Therefore, the best power allocation capacity from CHSO is clear.

\subsection{Power Allocation under Imperfect Channel Conditions}\label{Section_PE_PA_nPCC}
In order to evaluate the CHSO and HSO effectiveness in terms of optimal power allocation, three analysis for channel conditions were carried out: a) non-perfect monitoring of the OPMs, in section~\ref{OPM_results}; b) channel ageing effects, in section~\ref{ageing_results}; c) power instability, in section~\ref{dynamical_results}. The general parameters values adopted for both algorithms are described Table~\ref{tab:5}, while input parameters are depicted in Table~\ref{tab:9}, with the choice of $r_0=5.8318\cdot 10^{-6}$ (CHSO) and $r_0= 6.1873\cdot 10^{-7}$ (HSO).

\subsubsection{Non-perfect monitoring of the OPMs}\label{OPM_results}
there is an inaccuracy in the monitoring of the OPMs. Here, it is considered as a random variable $\epsilon_i \sim \mathcal{LN}\displaystyle (\mu, \sigma)$, where $\mu=0$ dB and $\sigma=0.16$ dB. These monitoring uncertainties corresponds to a maximum error $\epsilon_{i_{\max}}=0.6$ dB with high probability ($>0.9995$), commonly adopted in the optical networks considering inaccuracies from the OPMs~\cite{ReF:31,ReF:32,ReF:33}. This error is added into $i$th SNR during the power allocation procedure. Moreover, the adopted scenario assumes an operation at the BoL without power instability.

Fig.~\ref{fig:8}.b) depicts the velocity and the tendency of convergence, as well as the quality of the solutions. As can be observed, there is a decrease in the ${\rm NMSE}$ with the increase in the number of iterations. It is noticed that for early iterations the CHSO achieves better convergence performance when compared to HSO. In terms of convergence velocity, the CHSO (at $n=42$) is able to attain a NMSE $=  3.2\cdot 10^{-2}$ approximately  three times faster than HSO ($n=123$). On the other hand, similar NMSE values are found in the later iterations, {\it i.e}, $n\geq 125$ iteration, where both algorithms achieve an asymptotic NMSE $\approx 3.21 \cdot 10^{-2}$. Those results are affected by the OPMs inaccuracies. Indeed, comparing both algorithms performance operating under perfect monitoring condition, Fig.~\ref{fig:8}.a), the same asymptotic ${\rm NMSE}$ value has not been observed in both schemes. In this ideal scenario, the maximum power penalty resulted in $p_{i_{\max}}^{\textsc{chso}} = 0.26042$ dB and $p_{i_{\max}}^{\textsc{hso}}=0.26037$ dB.

\subsubsection{Channel ageing effects}\label{ageing_results}
\begin{figure}[htbp!]
	\centering
	\includegraphics[trim={1mm 0mm 1mm 0mm},clip, width=1\linewidth]{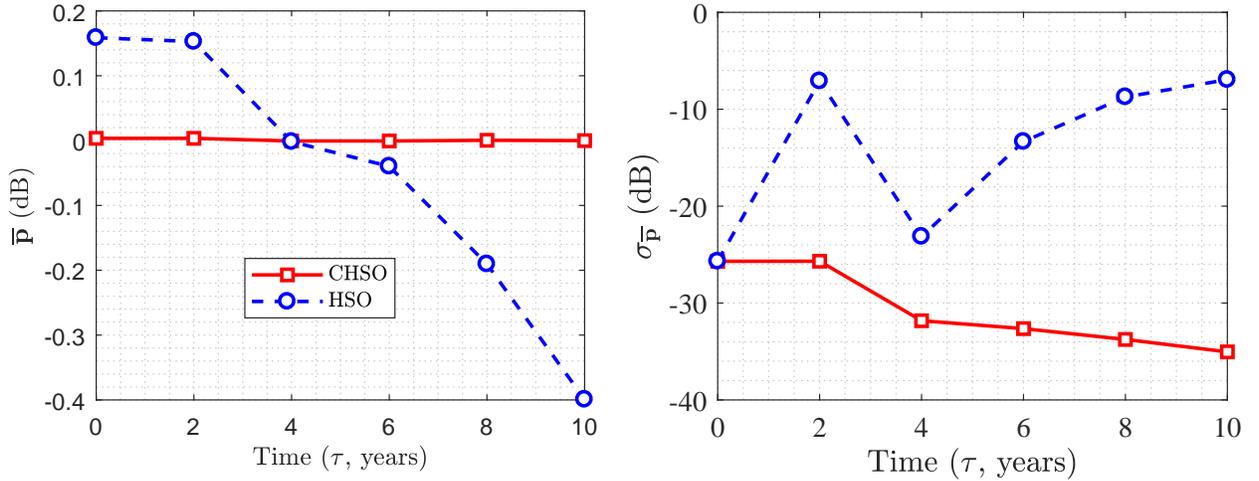}
	\caption{{({\bf a}) Expected value of power penalty for $M$-channels across time (at years), $\mathbb{E}{[\overline{\bf p}]}$, and ({\bf b}) their respective standard deviation ($\sigma_{\overline{\bf p}}$), for the CHSO and HSO.}}
	\label{fig:70}
\end{figure}

Under equipment ageing effects, Fig.~\ref{fig:70} proposes analyze the  power penalty trend against a multi-period incremental assuming $\tau=[0,2,4, \cdots, 10]$ years, representing the effect of ageing from BoL to EoL network. It illustrates the expected value of the power penalty from $M$-channels ($\mathbb{E}{[\overline{\bf p}]}$) across the time, as well as their respective standard deviation ($\sigma_{\overline{\bf p}}$). The ageing from the parameters is assumed as a linear function of time $\tau$. 

Elaborate further, it is possible to see in Fig.~\ref{fig:70}, that CHSO performs better when compared to the HSO.  $\mathbb{E}{[\overline{\bf p}]}$ and $\sigma_{\overline{\bf p}}$ are measured with the objective of evaluating the lower and upper bound of the power penalty of $M$-channels during EON lifetime. The upper and lower bound target are defined by $\varLambda_1$ and $\varLambda_2$, resulting in a  power penalty of $\bar{p}_i^{-}=-1.7407\cdot 10^{-2}$  [dB] and $\bar{p}_i^{+}= 4.3427\cdot 10^{-4}$ [dB] respectively. In other words, $\Psi_{\rm min}=\Psi^*-\varLambda_1$ is adopted as the minimum {RM} necessary to found the $\text{BER}^*$, whille $\Psi_{\rm max}=\Psi^*+\varLambda_2$ is adopted as the maximum {RM} to found the $\text{BER}^*$. In case of the {maximum RM}, CHSO is better than HSO, achieving a maximum value of $\mathbb{E}{[\overline{\bf p}]} \approx$ $3.4730\cdot 10^{-3}$ dB at $\tau=2$ against $\mathbb{E}{[\overline{\bf p}]} \approx$ $1.5863\cdot 10^{-1}$ dB at $\tau=0$. In terms of {minimum RM}, the CHSO found BER$^*$ all the time, a consequence of $\mathbb{E}{[\overline{\bf p}]}\approx$ $-6.0805\cdot 10^{-4}$ dB {$\leq \Psi_{\rm min}$} at $\tau=6$, while the HSO does not found BER$^*$, a consequence of $\mathbb{E}{[\overline{\bf p}]}\approx$ $-3.9924\cdot 10^{-1}$ dB {$\geq \Psi_{\rm min}$} at $\tau=10$. Therefore, the CHSO and HSO resulted at a margin  increasing of $6.0805\cdot 10^{-4}$ dB and $3.9924\cdot 10^{-1}$ dB, respectively, and presented a better saving energy. Besides, the $\sigma_{\overline{\bf p}}$ values found demonstrated that CHSO is more stable than HSO in terms of minimum energy expenditure to achieve the BER{$^*$}. In this context, CHSO is effective to mitigate the channel ageing effects.

\subsubsection{Power Instability}\label{dynamical_results}
Assuming now a dynamic scenario characterized by power instability or perturbation,  which can represent dropping or adding channels to the EON. After node add-drop channels an undesired effect reaches the surviving channels, herein modeled as a sine function in eq.~\eqref{eq:power_inst}, where $A_{\rm pert}=0.8$ dB and $f$=0.5 Hz represents overshoot and undershoot maximum adopted in the project of EDFA compensation of $\pm 1$ dB. Theses values assured the drops of the two routes, simultaneously~\cite{ReF:35}.

\begin{figure}[htbp!]
	\centering
	\includegraphics[trim={5mm 0mm 0mm 0mm},clip, width=.85\linewidth]{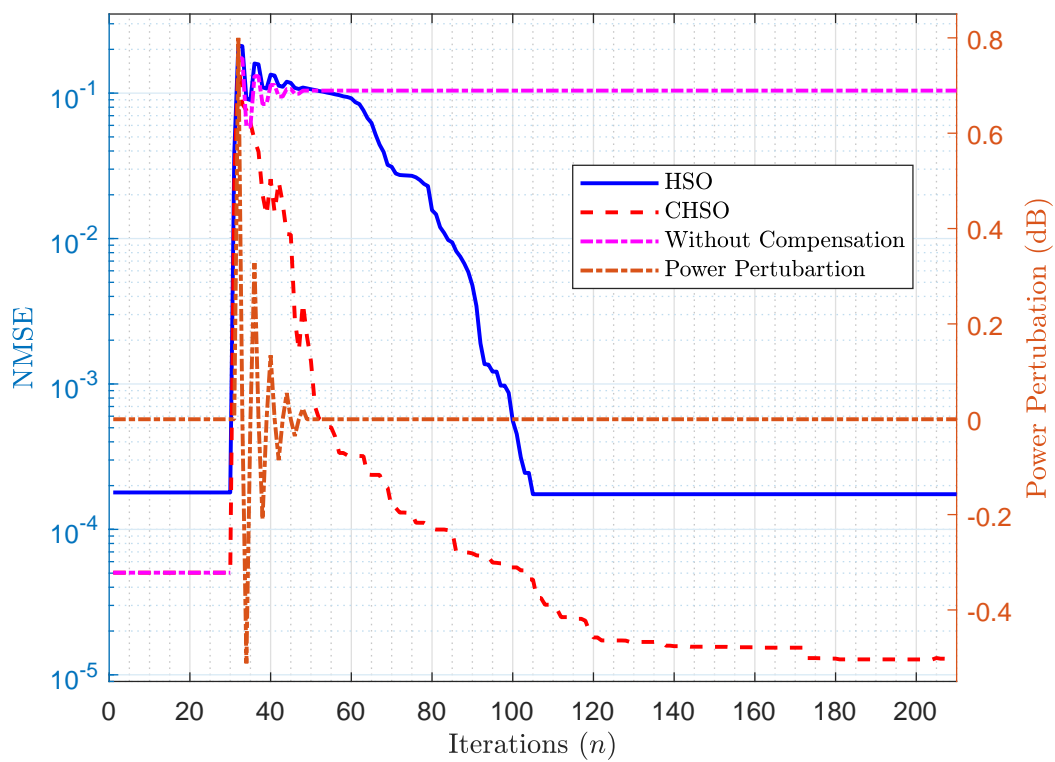}
	\caption{${\rm NMSE}$ for a dynamic scenario characterized by a power perturbation (${\rm pert}$) occuring in between $30<n\leq 49$ iterations. Two channels are dropped, ${\mathcal{R}}_{10}$ and ${\mathcal{R}}_{11}$, and three situation are taken: without compensation and compensation via CHSO and HSO.}
	\label{fig:11}
\end{figure}

In simulations of Fig.~\ref{fig:11}, a dynamic scenario has been modeled assuming a network optimized to operate with 12 users, such as in Table~\ref{tab:3} and Fig.~\ref{fig:3}. Thus, a fast variation is introduced at the node 8, where {${\mathcal{R}}_{10}$} and {${\mathcal{R}}_{11}$} are dropped at the iteration 30. This dropping results in four surviving channels ({${\mathcal{R}}_{4}$}, {${\mathcal{R}}_{8}$}, {${\mathcal{R}}_{9}$} and {${\mathcal{R}}_{12}$}) forward. These channels are affected by power fluctuations from node 8 to $\mathcal{D}$. The interval of perturbation occurs at $30<n\leq 49$. 

Elaborating furhter, Fig.~\ref{fig:11} illustrates the effect of the power perturbation under three situations: with and without compensation from CHSO and HSO. In case of no compensation, the launch power is assumed as optimized from the CHSO, and power adjustment is not carried out after the drop of the two channels. In other words, the channels ${\mathcal{R}}_{4}$, ${\mathcal{R}}_{8}$, ${\mathcal{R}}_{9}$ and ${\mathcal{R}}_{12}$ are penalized and theirs transmission power are not re-optimized, resulting in a NMSE$[210]=1.0406 \cdot 10^{-1}$. However, performance improvment can be attained deploying compensation in HSO and CHSO, resulting in a NMSE$[210] =  1.7455\cdot 10^{-4}$ and $1.2874 \cdot 10^{-5}$, respectively. It is evident the CHSO ability to escape from local minimum around $n=100$, as well as the behavior of both algorithms in the sense of following the power perturbation and in achieving the optimal power in latter iterations. 

A comparison between the initial and final NMSE value showed that for the HSO, similar values are found, {\it i.e.}, NMSE${_{1}}=1.7946\cdot 10^{-4}$ and NMSE$[1]=1.7455\cdot 10^{-4}$; and for the CHSO, a better final value is found, {\it i.e.}, a gap of $\Delta$NMSE$=3.76\cdot 10^{-5}$. 
Therefore, the power allocation assuming fluctuation from drop channels is validated and a better performance is found by the CHSO.

\subsection{Complexity} \label{Section_CC}
The computational complexity is evaluated in terms of mathematical operations and number of channels. In asymptotic terms, the HSO and CHSO have complexity of order of $\mathcal{O} (M^2)$. On the other hand, the complexity of GD algorithm is of order of $\mathcal{O} (M^3)$, as described in section~\ref{Complexity}. Aiming at attaining more accuracy in the complexity analyses, we have considered the mathematical operations from eqs.~\eqref{C_HSO},~\eqref{C_CHSO} and~\eqref{C_GD}. Three different system loadings have been adopted: {\bf A} has 12 channels (2,2 Tbps), as described in Table II and Fig.~\ref{fig:3};  {\bf B} has 120 channels (22 Tbps);  and  {\bf C}  has 240 channels (44 Tbps). {\bf B}  and  {\bf C} have the same topology of {\bf A}, {however theirs routes result of 10 and 20 times of {\bf A} ($\mathcal{R}_1, \cdots, \mathcal{R}_{12}$), respectively}. Those scenarios assume perfect channel conditions: operation at the BoL, static operation, and perfect monitoring of channel.

\begin{figure}[htbp!]
	\centering
	\includegraphics[trim={5mm 0mm 10mm 0mm},clip, width=.8\linewidth]{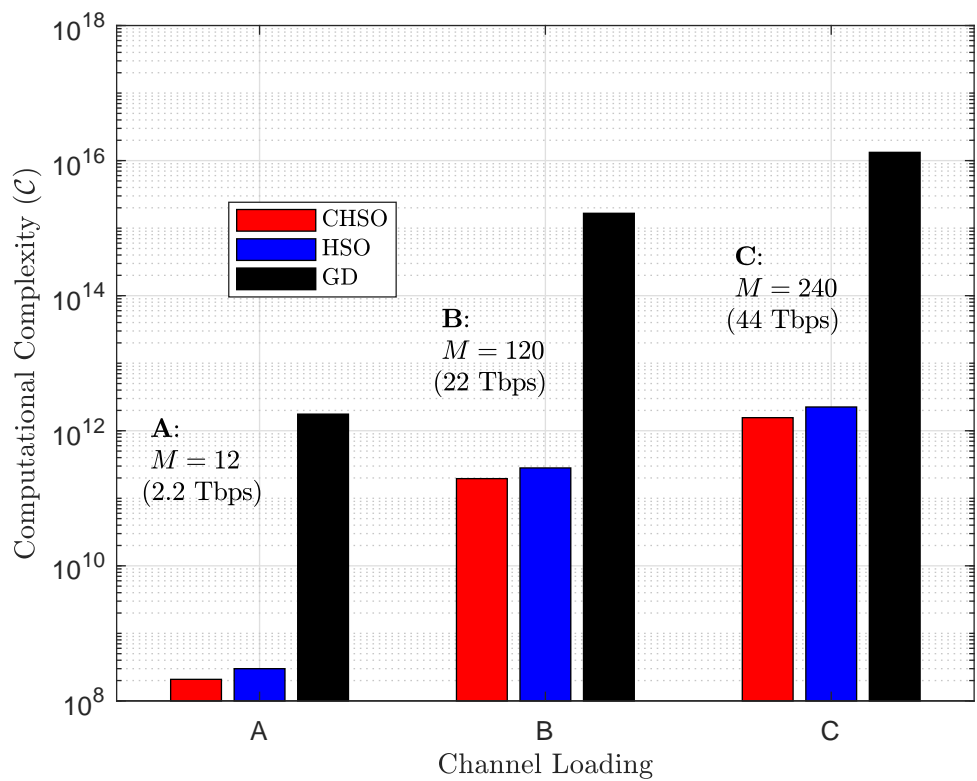}
	\caption{Complexity computational for three different channel  scenarios: CHSO, HSO and GD optical power allocation methods.}\label{fig:12}
\end{figure}

Fig.~\ref{fig:12} depicts the averaged computational complexity for the three algorithms operating under  {\bf A},  {\bf B} and {\bf C} scenarios. It also assumes optimized parameters from the Table~\ref{tab:9}. Those parameters result the worst-case for the computational complexity, {i.e.}, $K$ and $N_f$ can be reduced due to the increasing of $M$-channels and $r_0$, while $\omega$ can be reduced due to the increasing of non-linear effect. The CHSO has resulted in lower complexity than two methods. In addition,~{the computational complexity can be reduced by considering re-optimization of input parameters for any network operating conditions}. 	

\section{Conclusions}
The CHSO method proved to be a promising technique to resource allocation in elastic optical networks, especially by Nyquist wavelength division multiplexing (WDM) super-channels, combining competitive convergence speed, control capacity, non-linear effects mitigation, higher probability of success in lower iterations and lower penalties. The CHSO has demonstrated a higher ability to escape of local minimum caused by non-linear effects in scenarios where higher bit rates are required. The optimized parameters presented robustness considering conditional probability of success. Moreover, it resulted in a computational complexity in the order of $\mathcal{O} (M^2)$, much lower than the gradient descent method (of order of $\mathcal{O} (M^3)$), and marginally lower compared to the conventional HSO.

The conventional HSO has presented inferior performance regarding the CHSO. In terms of the optimization of parameters, a narrow conditional probability of success was found, resulting in a low ability for absorption of ageing effects and vast-variations, a consequence of higher sensibility to the parameters variation. Moreover, it was found worse penalties and lower convergence speed in case of dynamic scenarios. 

The CHSO performs power allocation in EONs with better performance-complexity tradeoff regarding both the HSO and the analytical GD method, considering non-perfect monitoring of OPMs, channel ageing effects and  dynamic scenario, that are the main realistic conditions from EONs operations. Such advantages result in a better margin reduction, energy efficiency improvement, and cost limitations.  In summary, inserting chaotic map procedure into the HSO (or CHSO) brought better performance-complexity balancing tradeoff.



\end{document}